\documentclass{aa}
\usepackage{psfig}
\begin{document}

%\thesaurus{03	 % Galaxies
%	   (03.0.1; % Emission line galaxies,
%	    03.1.1) % luminosity functions }

\title{Star forming rates between $z=0.25$ and $z=1.2$ from the 
CADIS emission line survey. }

\author{H. Hippelein\inst{1} \and C. Maier\inst{1} \and K. Meisenheimer\inst{1} \and C. Wolf\inst{1,2} 
    \and J.W. Fried\inst{1} \and B. von Kuhlmann\inst{1} \and M. K\"ummel\inst{1} \and S. Phleps\inst{1}
    \and H.-J. R\"oser\inst{1} }

\institute{ Max--Planck--Institut f\"ur Astronomie, K\"onigstuhl 17, 
            D-69117 Heidelberg, Germany
       \and Department of Physics, Denys Wilkinson Bldg., University of Oxford, 
            Keble Road, Oxford, OX1 3RH, U.K. }

\date{Received  2002; Accepted }

%\offprints{H. Hippelein}

\abstract{
The emission line survey within the Calar Alto Deep Imaging Survey
(CADIS) detects emission line galaxies by a scan with an imaging
Fabry-Perot interferometer. It covers 5 fields of $> 100\,\sq \arcmin$
each in three wavelengths windows centered on $\lambda \simeq 700$,
820, and 920\,nm, and reaches to a typical limiting line flux of $3
\times 10^{-20}$\,W\,m$^{-2}$. This is the deepest emission line survey covering
a field of several 100 $\sq \arcmin$. Galaxies between $z = 0.25$ and $z
= 1.4$ are detected by prominent emission lines (from H$\alpha$ to
[O\,{\sc ii}]372.7) falling into the FP scans. Additional observations with a
dozen medium band filters allow to establish the line identification
and thus the redshift of the galaxies to better than $\sigma_{\rm z} =
0.001$. On the basis of a total of more than 400 emission line galaxies 
detected in H$\alpha$ (92 galaxies), [O\,{\sc iii}]500.7 (124 galaxies), 
or [O\,{\sc ii}]372.7 (222 galaxies) we measure the instantaneous star 
formation rate (SFR) in the range
$0.24 < z < 1.21$. With this purely emission line selected sample we
are able to reach much fainter emission line galaxies than previous,
continuum-selected samples. Thus completeness corrections are much
less important. Although the relative [O\,{\sc iii}] emission line 
strength depends on excitation and metallicity and shows strong variation, 
the mean line ratios yield SFR[O\,{\sc iii}] values consistent with the 
SFR evolution. 
Our results substantiates the indications from previous studies (based 
on small galaxy samples) that the SFR decreases by a factor of $\sim20$ 
between $z = 1.2$ and today. In fact, for a 
$\Omega_{\rm m} = 0.3, \Omega_\lambda = 0.7$ cosmology, we find an exponential 
decline $\dot{\rho}_{\rm SFR} \propto \exp(-t_{\rm lookback} / 2.6$\,Gyr). 
This decrease of the SFR with time follows an exponential law which is 
compatible with the decreasing galaxy merger rate as expected from model 
calculations. 
The inferred SF density is in perfect agreement with that
deduced from the FIR emission of optically selected galaxies which 
is explained by a large overlap between both populations. We show that 
self-consistent extinction corrections of both our emission lines and 
the UV continua lead to consistent results for the SF density. 
}
\titlerunning{Star forming rates between $z=0.25$ and $z=1.2$}
\authorrunning{Hippelein et al.}
\maketitle

\section{Introduction}  %1

A number of studies of the star forming rate (SFR) were performed at different 
redshift regimes between the present epoch (Gallego et al. 1995) over intermediate 
redshifts (Tresse \& Maddox 1998; Sullivan et al. (2000, 2001); Lilly et al. 1996; 
Cowie et al. 1997; Connolly et al. 1997) up to redshifts $\sim3$ (Madau et al. 
1996; Pettini et al. 1998). Their results suggest, that the SFR has declined 
from a maximum plateau at $z\sim1.5$ to an one order of magnitude lower value in 
the present time. 

The most direct way to determine the SFR is to observe the luminosity density of 
the H$\alpha$ line, 
which is little affected by extinction, excitation, and metallicity, and for which 
Kennicutt (1983, 1992) has derived a well established calibration relation. 
For redshifts higher than 0.4, however, it is necessary to use emission lines 
other than H$\alpha$, the luminosity of which are not so clearly related to SFR 
(Cowie et al. 1997; Hogg et al. 1998; Rosa-Gonzalez et al. 2002), or to observe in 
the near infrared (Yan et al. 1999; Teplitz et al. 2000; Pettini et al. 2001). 
Observations in the far infrared generally led to higher rates (Rowan-Robinson 
et al. 1997; Hughes et al. 1998; Flores et al. 1999) 

As shown by Lilly et al. (1996), the star formation can also be derived from the 
UV luminosity of galaxies. Here however, the conversion to the SFR is severely 
affected by internal extinction, and for several years there was a huge gap 
between  UV determined rates and the rates derived by other observational methods, 
since the proper extinction correction was not known. Recent studies of the 
reddening in starburst galaxies (Calzetti et al. 1994, Jansen et al. 2001) have 
succeeded to close this gap by providing more reliable correction factors. 

Studies to relate the Balmer line luminosities to those of other emission lines 
and to the UV have been performed by Bell \& Kennicutt (2001) 
for the local universe, by Sullivan et al. (2000) for $z\sim0.15$, by Glazebrook 
et al. (1999) for $z=0.9$, and recently by Rosa-Gonzalez et al. (2002). 

In the study presented here a new observational technique is applied to derive 
star forming rates and their evolution in the redshift range 0.25 to 1.2: 
The emission line luminosities for medium redshift galaxies are measured 
photometrically on deep narrow band images.

The search and classification of emission line galaxies is based on both 
the multi-color and the Fabry-Perot observations of the Calar Alto Deep Imaging 
Survey (CADIS). 
While in previous determinations of the star forming rates from emission line 
luminosity densities the object samples were selected via broad band magnitudes, 
in the present study it is based on the strength of the emission lines directly. 
Since spectrophotometric observations reach fainter limiting magnitudes than 
spectroscopic ones, the CADIS survey can reach galaxies with very low continuum 
brightnesses down to the extreme case of galaxies which are only detectable by 
their emission lines. In addition, the multi-color survey also provides UV 
luminosities, which allow within one and the same data base the comparison of the 
emission line luminosity densities with those in the UV. 

The strategy of this deep imaging survey is to perform narrow band observations of 
relatively large fields in order to detect emission line objects down to very low 
line fluxes. In addition 15 broad and medium images are observed which are used 
to detect secondary emission lines and to derive spectral energy distributions 
for all detectable objects. These data are used to determine the spectral types 
and redshifts for a large number of galaxies by means of a multi-color 
classification routine. 

Throughout this paper we use $\Omega_{\rm m}=0.3$, $\Omega_{\Lambda}=0.7$; 
$H_{\rm 0}=70$\,km\,sec$^{-1}$\,Mpc$^{-1}$ if not parameterized by $h$.

\section{Observations and data handling}  %2

\subsection{Observations}  %2.1

The observational principle of the CADIS survey is described in Meisenheimer 
et al. (2002). In the present paper we will only address in detail the observational 
and reduction procedures relevant to the study of emission line galaxies. 

The observations are performed on a number of fields distributed over the 
sky selected at areas of minimum IRAS 100\,$\mu$m emission (low galactic extinction) 
and free of bright stars. 

In order to detect emission line objects, narrow band observations were 
done in three wavelength windows, which are relatively free from OH night-sky 
emission. 
In each of these windows observations are performed with a scanning Fabry-Perot 
interferometer (FPI) at 8 or 9 equally spaced wavelength steps, thus providing 
a spectral line scan (henceforth called FP scan). 
The three wavelength windows are specified by A, B, and C, and are located at: 
\medskip

\indent window ~ FP-A : ~ ~ $\lambda = 702\pm6$\,nm, \\
\indent window ~ FP-B : ~ ~ $\lambda = 820^{+4.5}_{-6}$\,nm, \\
\indent window ~ FP-C : ~ ~ $\lambda = 918\pm8$\,nm. 
\medskip

For selecting the correct interference order of the FPI, pre-filters with FWHMs 
of $\sim20$\,nm are used.

The FP observations in the B window were carried out with the CAFOS focal reducer at 
the 2.2\,m telescope on Calar Alto, Spain; those in the A and C windows were carried 
out with the 
MOSCA focal reducer installed at the 3.5\,m telescope. Both focal reducers are 
equipped with Fabry-Perot (Queensgate) etalons in the parallel beam, with free 
apertures of 50 and 70\,mm, and with air gaps of $\sim$8\,$\mu$m. By choosing 
the appropriate interference order, the spectral resolving power for both etalons 
was set, to  $\sim$750\,km\,s$^{-1}$. The actual spectral resolutions used were 
1.8, 2.0, and 2.4\,nm. 
Exposures were also done with the pre-filters alone in order to measure the 
continuum level underneath the emission lines found in the FP scans. 
At the 2.2\,m telescope a flux limit of $F_{\rm lim} = 3 \times 10^{-20}$\,W\,m$^{-2}$ 
was achieved after 3 hours of integration, sufficient to detect line emission 
of a galaxy with a star forming rate of $\sim$1.0\,M$_{\odot}$\,y$^{-1}$ 
at redshift z=1.2.

\begin{table*}[hbt!]
\caption{Emission lines and corresponding (veto) filters for measuring their fluxes; 
filter specifications are noted in terms of central wavelength / full width at half 
maximum, both given in nm. }
\begin{center}
\begin{tabular}
{     c   |    c    |    c    |   c         c       c        c        c    }
 \hline   
 \hline   
          &         &         &        &        &        &        &        \\[-3mm]
redshift  &   FP    &   FP    & \multicolumn{5}{c}{Filter used to observe} \\
 bin      &  line   & Window &H$\alpha$& [OIII] &H$\beta$&H$\gamma$& [OII] \\
          &         &         &        &        &        &        &        \\[-2mm]
 \hline   &         &         &        &        &        &        &        \\[-3mm] 
 0.25     &H$\alpha$& 820\,nm &  FP-B  & 628/17 & 611/16 & 535/14 & 465/9  \\
 0.40     &H$\alpha$& 918\,nm &  FP-C  & 702/18 &   -    & 611/16 & 522/16 \\[-1mm]
          &         &         &        & + FP-A &        &        &        \\
 0.40     & [OIII]  & 702\,nm & 909/30 &  FP-A  &   -    & 611/16 & 522/16 \\[-1mm]
          &         &         & + FP-C &        &        &        &        \\
 0.64     & [OIII]  & 820\,nm &   -    &  FP-B  &   -    &   -    & 611/16 \\
 0.88     & [OII]   & 702\,nm &   -    &   -    &   -    & 815/25 &  FP-A  \\
 1.20     & [OII]   & 820\,nm &   -    &   -    &   -    &   -    &  FP-B  \\[-3mm]
          &         &         &        &        &        &        &        \\
 \hline
\end{tabular}
\end{center}
\end{table*}

Since this observational method bears no color selection effect a priori, one expects 
to find also emission line galaxies with very faint continua, such as distant H\,{\sc ii} 
galaxies and primeval galaxies at high redshift, detectable only in the light of their 
Lyman-$\alpha$ lines. 

In order to decide which emission line was detected in the FP scan, we placed 
for each FPI band a set of medium band filters at wavelengths where secondary 
prominent lines would show up (Fig.\,1). Since these filters allow to 
exclude certain line classifications - {\it e.g.,} the presence of a line in any of these 
filters excludes that the line detected in the FP scan is [O\,{\sc ii}] or Ly$\alpha$ -, 
they are named {\it veto} filters. These veto filters (see Table\,1) are included 
in a the set of 15 medium and broad band filters, which range from 400 to 2200\,nm in 
wavelength and provide additional information about the spectral energy distribution 
(SED) of the line emitter and its spectral type. Table 1 lists for the redshift bins 
discussed in the present paper, the filters where the secondary lines are expected.

\subsection{Data base}             %  2.2

The present study is based on the data sets from four fields (size $\sim10\times10$\,arcmin) 
of the CADIS emission line survey which are completely reduced up to now, as 
listed in Table 2. 
In these fields we selected the emission line galaxies detected 
in the light of their prominent emission lines H$\alpha$, [O\,{\sc iii}]$\lambda500.7$, 
or [O\,{\sc ii}]. 

For the window B around 820\,nm, the data of the 01h, 09h, and 16h fields were used to derive 
luminosity functions and star forming rates at redshifts of $z=0.25$ (H$\alpha$ line), 
0.64 ([O\,{\sc iii}] line), and 1.2 ([O\,{\sc ii}] line). 
For the window A, around 702\,nm, 
the data of the 09h, 16h, and 23h fields were used. Here the redshifts studied 
are $z=0.40$ ([O\,{\sc iii}] line), and $z=0.88$ ([O\,{\sc ii}] line). The H$\alpha$ galaxies 
appearing at redshift 0.08 are too rare to provide a useful database. 
The data of window C were only used for consistency checks, but not for deriving 
luminosity functions and star forming rates, since the classification for this 
window is not yet fully understood.

\begin{table}[ht]
\caption{Fields observed with CADIS used in the present paper.}
\begin{center}
\begin{tabular}
{     c   |    c     |           c                  c         |     c     }
 \hline   
 \hline
          &          &                      &                 &          \\[-3mm] 
   Field  &    FP    & ~ ~ ~ ~ ~ ~ Center   &coord. ~ ~ ~ ~ &   Area   \\
  name    &  window  &      $\alpha$(2000)  & $\delta$(2000)  & arcmin$^2$\\
          &          &                      &                 &          \\[-2mm]
 \hline
          &          &                      &                 &          \\[-3mm] 
   01h    &   B ~ ~  & 01$^h$47$^m$33.3$^s$ &   02 19 55''    &    103   \\
   09h    &   A, B   & 09$^h$13$^m$47.5$^s$ &   46 14 20''    &    100   \\
   16h    &   A, B   & 16$^h$24$^m$32.3$^s$ &   55 44 32''    &    106   \\
   23h    &   A ~ ~  & 23$^h$15$^m$46.9$^s$ &   11 27 00''    &     98   \\[-3mm] 
          &          &                      &                 &          \\
 \hline
\end{tabular}
\end{center}
\end{table}

The number of galaxies found per field and redshift interval, with emission line 
fluxes above our detection limit is of order of 30, providing emission line 
galaxy samples large enough for statistical studies (see also Table 3). 

The wavelengths can be set with the FP etalon within an accuracy of $\pm0.2$\,nm. 
The wavelength calibration was done several times during the night using the calibration 
lamp lines Rb\,794.5\,nm and Ne\,692.95\,nm, and was stable within 0.15\,nm during 
the observations. This is roughly one order of magnitude less than the width of the 
instrumental profile of $\sim2.0$\,nm, which was adjusted by selecting the appropriate 
interference order of the FPI. 
Towards the edges of the fields, the transmitted 
wavelengths declined by 0.5\,nm, due to the increasing angle of incidence in the FP etalon. 
In combination with the dithering used for the observations, this leads to an additional 
wavelength uncertainty of $\pm0.1$\,nm in the edges of the frames. 
Including the error of the line fit (see Sect. 2.5), the total uncertainty for a medium 
strong line thus arises to about $\pm0.4$\,nm in $\lambda$, or about $\pm0.0008$ in redshift. 
A comparison of the redshift values for objects with $z\sim0.40$, which can be seen in 
both the window A (by their [O\,{\sc iii}] line) as well as in window C (by H$\alpha$) yields 
the same value for the redshift uncertainty.

\subsection{Data reduction}  %2.3

The data reduction includes subtraction of bias, flatfielding, and removal of 
cosmics and detector defects by overlaying dithered images and replacing the bad 
pixels by the kappa-sigma-clipped mean of pixel values in the other images. 

Special care has to be taken for the flatfielding, since internal reflections 
between the focal reducer optics and the Fabry-Perot etalon can give rise to 
straylight contributions in the central area of the flatfields, thus leading 
to a incorrect flux calibration of the FP data. In order to overcome this 
problem, flat field exposures are taken through a regularly spaced multi-hole mask, 
allowing to measure the instrument transmission without noticeable straylight 
contribution. 

After overlaying all images to the same world coordinate system, the images for 
each wavelength setting and filter are then added up. For each of these sum frames 
the object search engine SExtractor (Bertin \& Arnouts, 1996) is applied, with 
thresholds adjusted to their seeing and exposure depth. The object positions are 
corrected for the distortion of the camera optics, and the object lists are merged 
to a MASTER table with averaged positions. For merging all objects are considered to be 
identical which fall into a common error circle of 1\arcsec\ radius. 

The morphology parameters of an object are determined on the sum frame where the 
object shows the highest S/N ratio, using the photometry package MPIAPHOT 
(Meisenheimer \& R\"oser, 1993). 
The photometry is performed on each individual frame, by integrating the photons 
at the centroid object positions, with a Gaussian weight distribution, the width 
of which is determined such that the convolution of the seeing PSF with the weight 
function results in a common PSF for all frames. The obtained flux is then 
calibrated by tertiary spectroscopic standard stars established in each CADIS field. 
Finally, the single frame flux values are S/N-weight averaged with errors 
derived from the counting statistics. 
Since the photometry weight function is normalized to give correct fluxes for 
stellar objects, the fluxes of extended objects are underestimated and need 
a correction according to their morphological parameters.

\subsection{Selection of emission line galaxy candidates}  %2.4

The pre-selection of emission line galaxies is based on the fluxes in 
the FP images and in the pre-filter image. Candidates have to fulfill two 
criteria: (1) For at least one FP wavelength, the signal has to be larger 
than the upper limit of the noise distribution, typically located near 5$\sigma$. 
(2) The signal-to-noise ratio of the line feature 
in the FP scans above the pre-filter flux is higher than ~2.5, equivalent to 
about $2\times 10^{-20}$\,W\,m$^{-2}$ (here we choose a very low threshold 
in order to be complete; a much stricter selection is possible on the basis 
of the line fits described below). This pre-selection yields 
a few hundred of emission line galaxy candidates per field and wavelength window. 

Due to reflections within the Fabry-Perot etalons employed in CADIS, bright 
objects, mostly stars, are accompanied by features about 6 
magnitudes fainter than the father object, and appear at a fixed offset 
position (41\arcsec) away from it. 
Since in the pre-filter images no object should be seen at the same positions, 
these ghosts can be easily sorted out and rejected from the candidates list. 

Another type of ghosts on the single FP frames is produced by reflections of 
bright objects in between FP etalon and pre-filter. These ghosts appear 
symmetrically to the optical axes and, due to dithering, they show up at 
different positions in every frame. Thus these are easily identifiable 
and already strongly suppressed by the correction for cosmic rays in the 
standard data reduction. 

Spurious objects were found to occur in the wings of bright objects, due 
to variations in the extended wings of the point spread function. A 
scatter of the flux measured in the Fabry-Perot images at the position of a 
spurious object can pretend an emission line. 
In order to find and reject these artifacts, all candidates with a 
bright neighbour were flagged. Based on eye inspection, a simple 
criterion was derived, which separates real objects from artifacts rather 
successfully: If 

$m_{\rm R}({\rm next\,bright\,object}) + dist({\rm arcsec}) < m_{\rm crit}$, 

with $m_{\rm crit}=24.0$, the feature is considered as an artefact and removed 
from the list of candidates. It was verified that this procedure does not lead 
to a loss of any ``good'' objects.

\subsection{Analysis of emission line objects} %  2.5

After pre-selection, the emission line galaxy candidates are classified, 
by analysing their emission line features in the FP scans as well as their 
SEDs, determined by the multi-filter observations. 
For the classification, three criteria are used: the shape of the 
emission line observed with the FPI, the emission line spectrum, 
derived from medium band veto filters placed at wavelength 
where prominent secondary lines are expected, and the continuum SED.  

This analysis and classification procedure is performed in five steps, 
which are described in the following in detail. 

\paragraph{(1) Line Fit: }
First step is the analysis of the photometric FP data by line fitting. 
The width of the line profiles are fixed to the width of the instrumental profile 
(1.8, 2.0, and 2.4\,nm for the windows A, B, and C respectively) for the 
particular wavelength window. 

Line fluxes were derived from the areas under the fitted line curves. 
For the fit, a modified Gaussian function was chosen of the form 
\begin{center}
 $f(\lambda) = f_{\rm 0} \times exp\big\{-1.96 |(\lambda-\lambda_{\rm 0}/\Delta\lambda|^{1.5}\big\}$, 
\end{center}
where $\Delta\lambda$ is the instrumental resolution (FWHM). This function has a more 
pronounced peak and broader wings than the normal Gaussian profile, 
and is rather close to the shape of the instrumental profile of the FP etalon. 
The area under this line profile is $A = 1.12 f_{\rm 0} \times \Delta\lambda$. 
Accounting for the fact, that the wings of the real FP profile are somewhat 
broader than in the above function, we use $A = 1.25 f_{\rm 0} \times \Delta\lambda$. 
This factor was verified by a comparison of the fluxes of the [O\,{\sc iii}] lines 
from $z\sim0.4$ galaxies as determined in the FP scans (window A), with those 
determined from the excess flux in the medium band filter located at 
$\lambda=700$\,nm above the continuum. 

Since the detected line can be any emission line occurring in the observed 
wavelength window at the appropriate redshift (e.g., H$\alpha$ at 0.25, [O\,{\sc iii}] 
at 0.64, etc., for the B window at 820\,nm), the classification routine has 
to investigate the solutions for all bright emission lines. Thus, we consider 
a total of eight possible line identifications in our fitting procedure: 
[S\,{\sc ii}]\,673, H$\alpha$, [O\,{\sc iii}]\,501 and 496\,nm, H$\beta$, 
H$\gamma$, [O\,{\sc ii}], and Lyman-$\alpha$. The number of galaxies expected 
to be seen in other lines is negligible (see Sect. 3.3) and does not justify to 
include them into this list. 
The H$\alpha$ line fit includes the [N\,{\sc ii}] doublet. With the the spectral 
instrumental resolution of $\sim 2.0$\,nm it is possible to separate the 
[N\,{\sc ii}]\,658.3 line in about 50\% of the cases successfully from the H$\alpha$ 
line with [N\,{\sc ii}]658.3/H$\alpha$ line flux ratios in between 0.1 and 0.5. This is 
generally the case for $S/N(H\alpha) > 4.4$. For the ten strongest H$\alpha$ 
emitter the ratio is [N\,{\sc ii}]/H$\alpha = 0.21\pm0.05$. 
The maximum line flux ratio [N\,{\sc ii}]658.3/H$\alpha$ is limited to 0.5 in the fit. 

In order to make use of the pre-filter fluxes, the line flux estimated from 
this line fit is subtracted from the pre-filter flux allowing for the filter 
transmission at the specific wavelength of the line. 
The line corrected pre-filter flux - and its error - is included 
as an additional continuum data point for an improved line fit. 
Examples for fits to different lines are shown in the right panels of 
Fig.\,1, where the observed pre-filter fluxes are marked as solid bars with 
their errors, and the corrected pre-filter fluxes as dashed lines. 
Both values are also marked in the SED plots in the left panels.

\begin{figure*}[ht!]               % Fig 1
\centerline{\vbox{
\psfig{figure=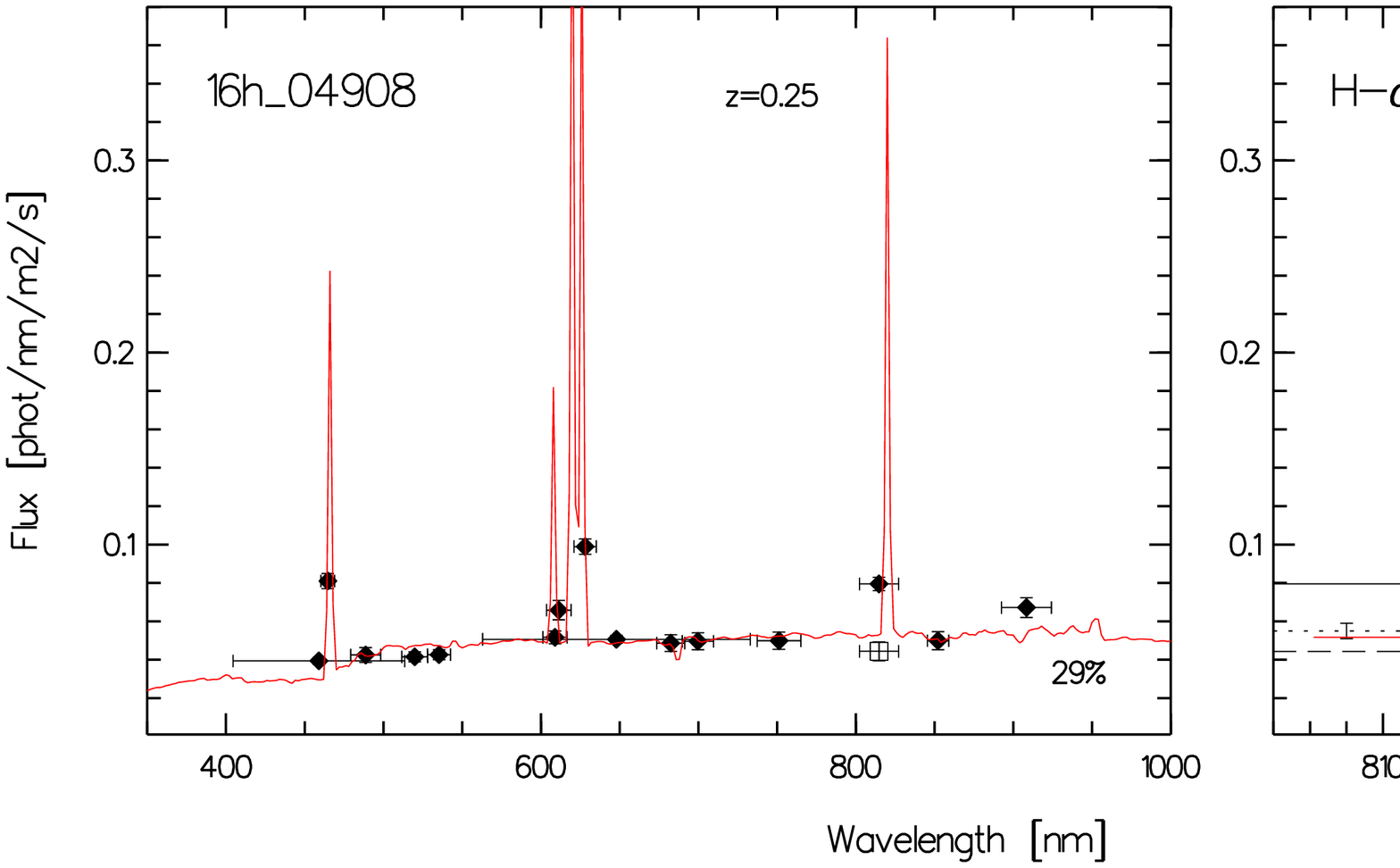,clip=t,width=12.8cm,height=4.48cm,angle=0}\vspace*{1mm}
\psfig{figure=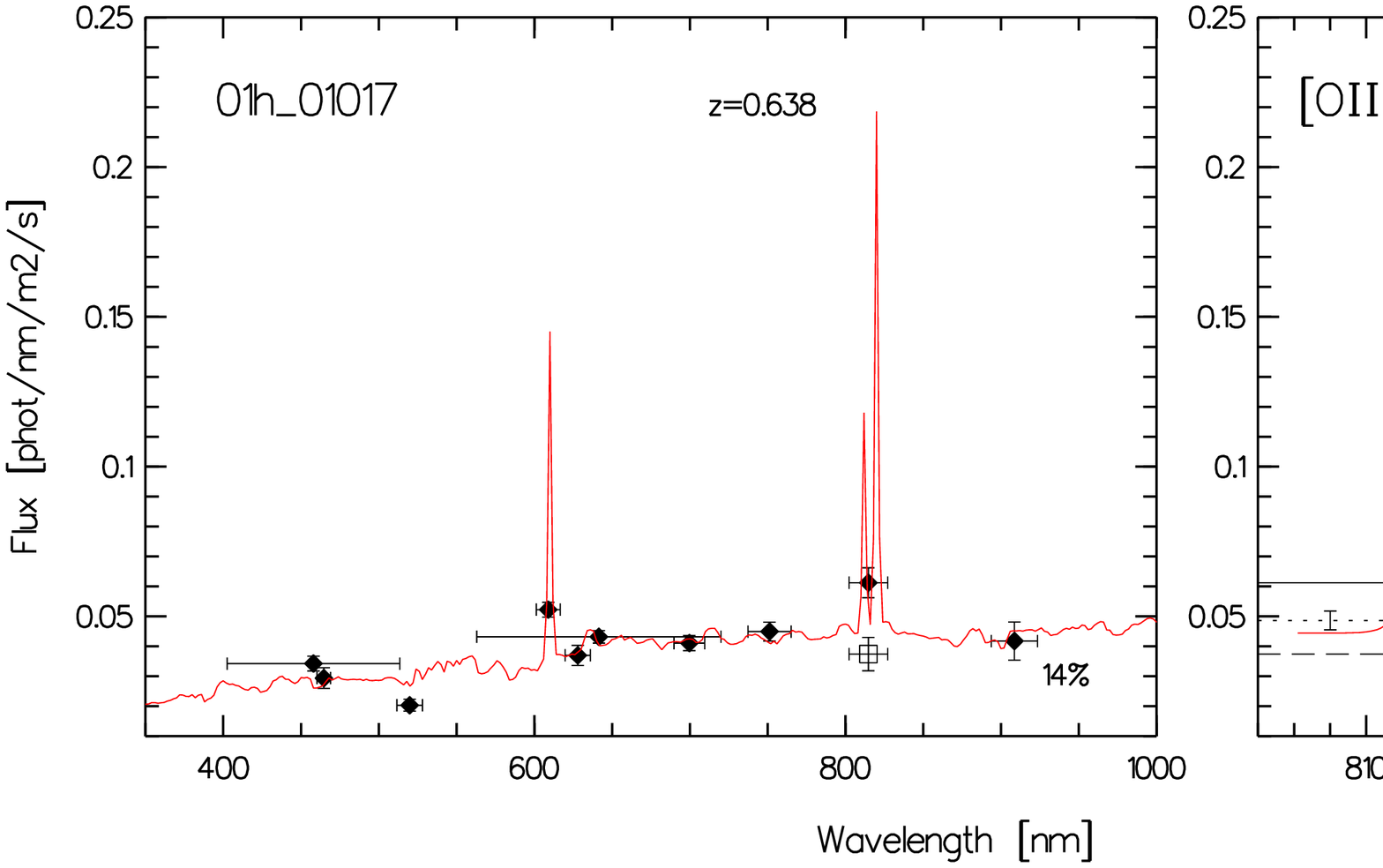,clip=t,width=12.8cm,height=4.48cm,angle=0}\vspace*{1mm}
\psfig{figure=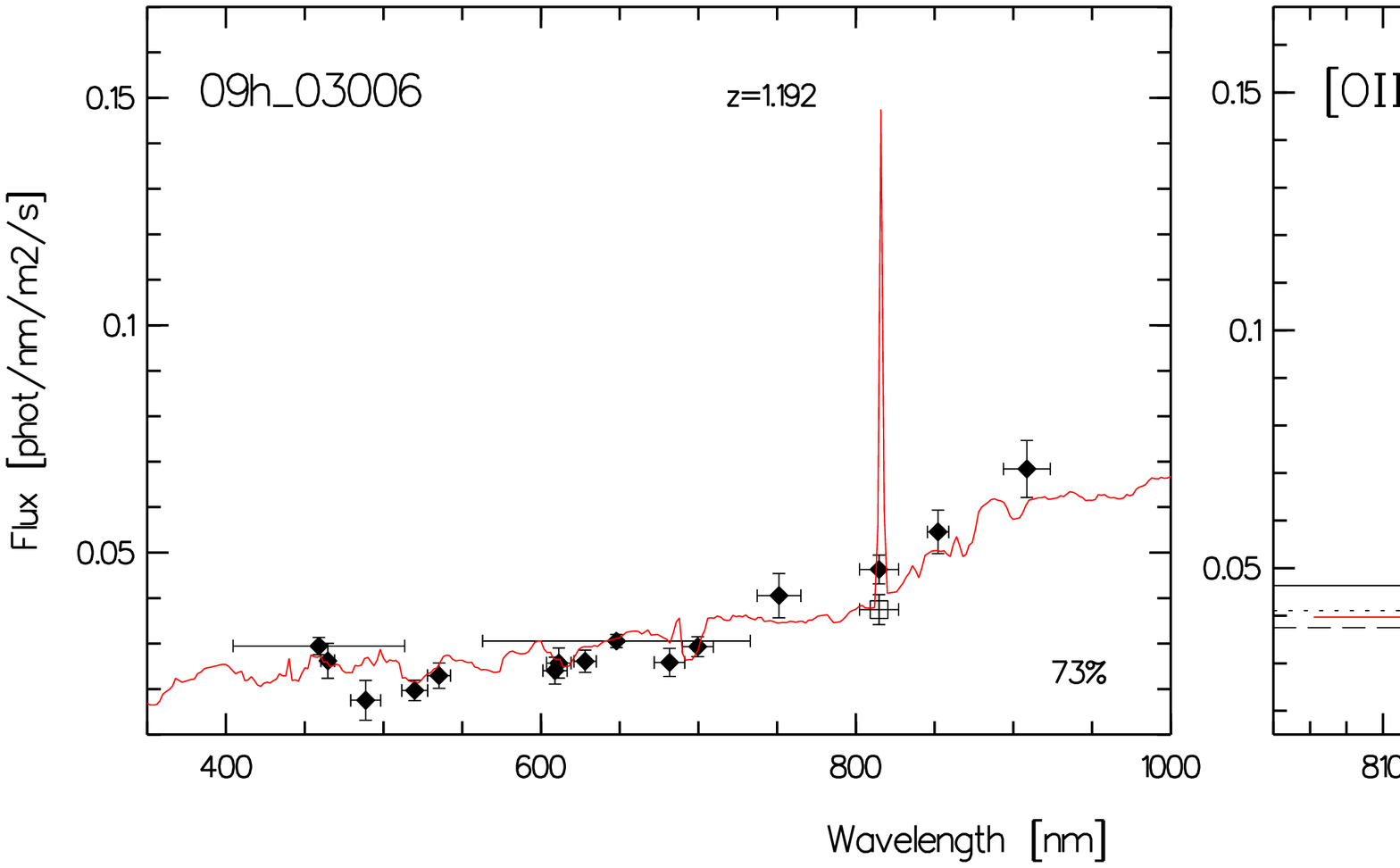,clip=t,width=12.8cm,height=4.48cm,angle=0}\vspace*{1mm}
\psfig{figure=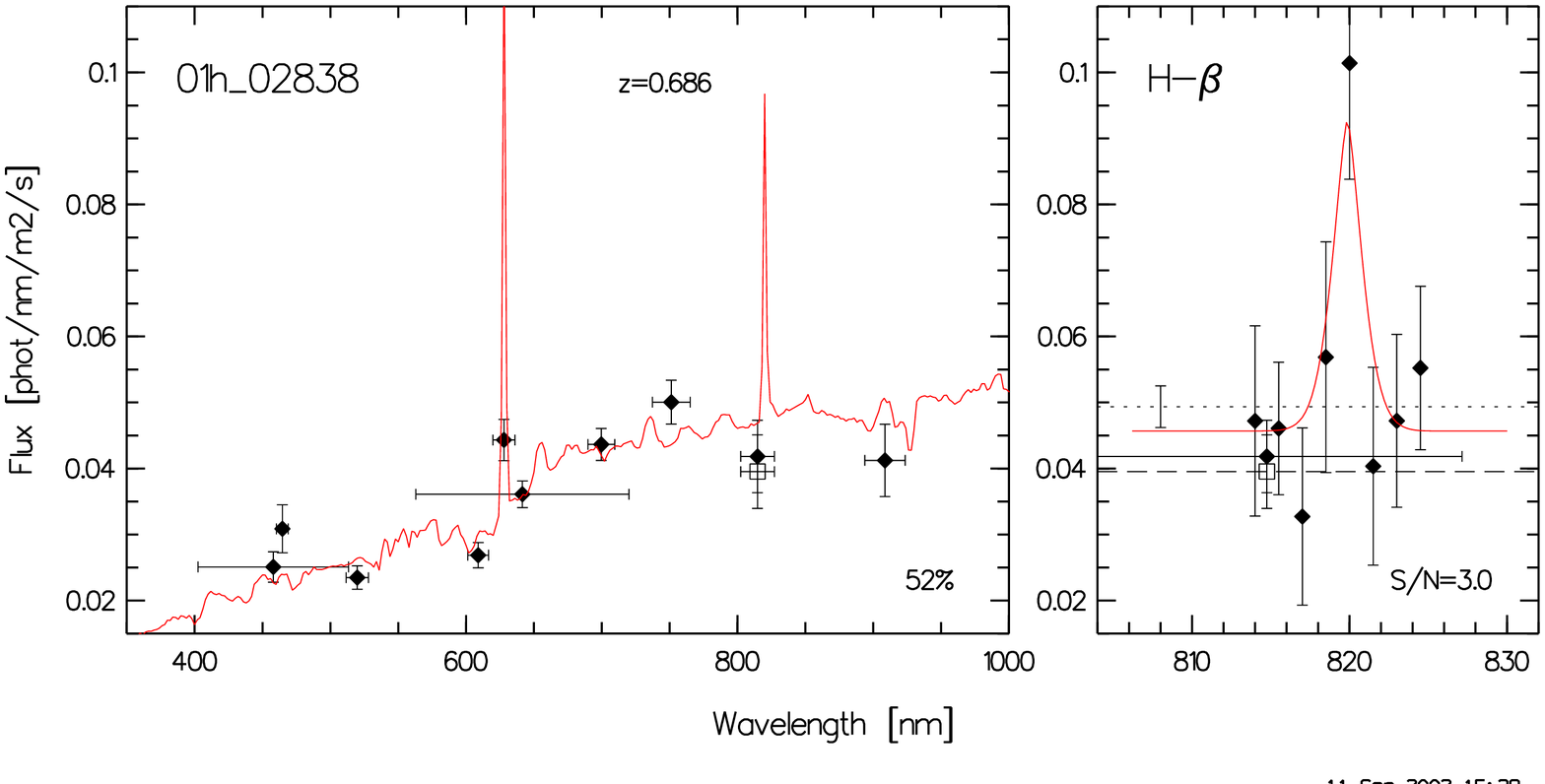,clip=t,width=12.8cm,height=5.23cm,angle=0}}}
\caption[]{Emission line fits (right panels), and smoothed continuum fits 
(left panels) for a $R = 22.9$\,mag galaxy at redshift 0.250 (top), a $R= 23.1$\,mag 
galaxy at $z=0.638$ (second row), and a $R = 23.4$\,mag galaxy at $z=1.192$ (third 
row). In the bottom panel, a $R= 23.3$\,mag galaxy at $z=0.688$ is shown; since 
there is no filter placed at 840\,nm, the strength of the [O\,{\sc iii}] lines 
is unknown for this case. The numbers in the lower right edges of the right panels 
specify the signal-to-noise ratios for the line fits. Filled circles represent 
observed data, empty squares the emission-line corrected pre-filter fluxes.} 
\end{figure*}

\paragraph{(2) Continuum SED fit: }
In the next step, for every of the eight redshifts determined above, 
template spectra published by Kinney et al. (1996) were fitted to the SED derived 
from the multi-filter photometry. This is done by means of colors, similar 
to the multi-color classification procedure described in Wolf et al. (2001), 
yielding a most probable template spectrum. In order to achieve better fits 
to the data, we allow for an extra reddening correction of the template spectra 
up to $E_{\rm B-V}$ of 0.3. The latter is important for distant samples which 
contain generally brighter and more massive galaxies with more reddening than 
the Kinney et al. templates (see sect. 3.5).

Since we want to determine the fluxes in the emission lines from the signal 
excesses in the veto filters above the continuum (see step 4), line free 
template spectra are needed. Thus, the Kinney-Calzetti template spectra were altered 
by cutting out the prominent emission lines. As a consequence, those filter data 
where secondary emission lines are expected, are not usable for the fit. Thus, 
the SED fit is purely based on the continuum part of the spectra. 

For the template spectrum which fits best to the observed SED a reduced $\chi^2$ 
is derived.

\paragraph{(3) Improved line fit: }
The fit to the continuum spectrum provides, for each of the eight 
possible line identifications, also a new continuum level at the wavelength of the FP window, 
which is marked in either of the line plots in Fig.\,1 as a dotted horizontal line. 
The error bar is estimated from the quality of the continuum fit, and is comparable 
to that of the pre-filter flux. 
Ideally, this continuum flux level should agree with the corrected pre-filter flux (step 1). 
A high discrepancy between the two values indicates that either the SED fit is bad 
(due to incorrect line classification), or the line feature seen in the FP scan 
is awkward. This information is used to produce another $\chi^2$ contributor, 
calculated from the discrepancy and the intrinsic errors of the two flux levels. 

Including the continuum level derived above, a new revised 
line fit is performed in the same way as in step 1. In this second iteration, however, 
bad line classifications become visible by large offsets between the continuum 
level given by the SED fit and the corrected pre-filter baseline, leading to a 
lower S/N ratio for the corresponding line fit. This also happens when 
the line feature in the FP data is in reality a fake detection due to a statistical 
fluctuation of the FP signals. 

The new S/N value is later used as critical indicator for the reality of the line 
feature seen in the FP scan and whether the object will be accepted as an emission 
line galaxy.

\paragraph{(4) Line spectrum fit: }
The continuum fit found this way serves as a baseline to determine the fluxes 
of the other (secondary) emission lines. 
For each of the eight possible line identifications, the fluxes for the secondary 
lines were derived from the difference between the flux measured in the 
`veto filters' (the filters where other prominent lines 
would be expected) and the continuum level predicted from the template fit. 
For the calculation of the line strength, the excess flux was normalized by 
the transmission of the filter at the wavelength given by the redshift derived 
from the FP observation. In cases where the veto filter included more than one 
emission line which generally happens for the [O\,{\sc iii}]\,501/496 doublet, the 
line flux ratio was fixed to the canonical values. 
Fig.\,1 shows such fits for several objects in the left panels. In all spectra, 
except for the [O\,{\sc ii}] galaxy, a number of `secondary' lines can be seen. 

The thus determined line flux ratios 
are compared with a catalog of observed line ratios for about 500 nearby 
galaxies based on data from the literature, including a range of Seyfert galaxies 
to compact dwarfs, most notably from French (1980), McCall et al. (1985), 
Popescu \& Hopp (2000), Veilleux \& Osterbrock (1987), Vogel et al. (1993). 

A $\chi^2$ is estimated in the following way: In the multidimensional 
space made up by all line ratio combination possible for the respective 
classification of the line observed in the FPI window, the mean 
distance of the considered galaxy to the three closest tabulated 
ratios was normalized by the error ellipse of the observed ratios. 
This method prevents a bias towards the most common line ratios, but allows 
also galaxies with extreme line ratios, as long as there are some objects 
with similar ratios tabulated. 
Again, a $\chi^2$ for the line ratios is estimated for each of the eight 
line cases.

\paragraph{(5) Finding the best solution: }
To decide between the possible line identifications an overall reduced $\chi^2$ 
is derived, which is the mean of the four $\chi^2$ values, of the SED fit (see 
step 2), of the line fit (step 3), of the discrepancy of the baselines (step 3), 
and of the line ratios (step 4). This mean $\chi^2$ is then converted into 
an overall probability for the line classification according to ~ $P = exp(- \chi^2/2)$. 
The case with highest probability is then selected as the most probable 
classification. 

Fig.\,1 depicts three examples of SED and emission line fits for galaxies 
at redshifts 0.250, 0.638, and 1.192 with strong emission lines, together 
with an example for galaxy with a line signal near the S/N cutoff (see 
Sect. 3.1), which was classified as H$\beta$ at $z=0.686$, and verified 
spectroscopically to be at $z=0.687$. In the plot, the fluxes are given in 
units of ~photons\,sec$^{-1}$\,nm$^{-1}$\,m$^{-2}$.

\subsection{Comparison with spectroscopic observations} % 2.6.

The decision, if an emission line galaxy candidate is considered as real or 
not is based on the S/N ratio of its line fit. To avoid contamination of the 
galaxy samples by fakes on the one hand and to be complete as possible on the 
other, an optimum S/N cutoff value has to be determined. 

Spectroscopic follow-up observations were carried out for a number of emission 
line galaxy candidates with the MOSCA spectrometer at the 3.5m Calar Alto telescope, 
with LRIS at Keck, and with FORS1 at VLT. 
We use the results of these observations to compare them with the predictions 
from our emission line classification, and to find the optimum S/N for a 
reliable classification. This comparison is depicted in Fig. 2.

\begin{figure}[ht!]               % Fig 2
\centerline{
\psfig{figure=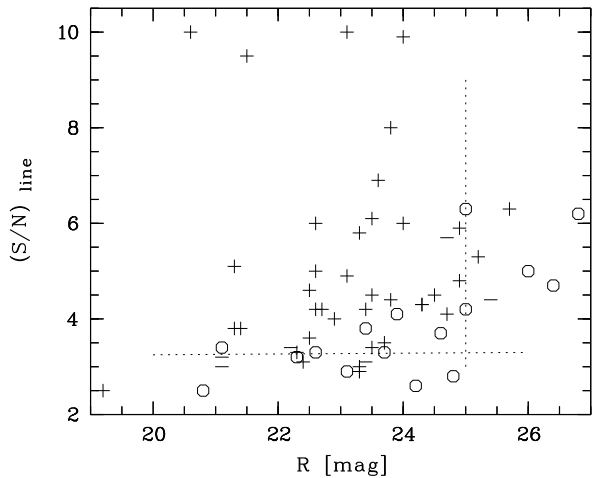,clip=t,width=8.0cm,angle=0}}
\caption[]{Correct and incorrect predicted line identifications for 
the spectroscopically observed galaxies, for varying $R$ magnitude and S/N ratio. 
+ stands for correct, - for incorrect identification. A circle indicates, that no 
line could be verified at the predicted wavelength.}
\end{figure}

The diagram shows, that for a S/N cutoff at 3.3, and for $R<25$\,mag (dotted 
lines), the probability for a correct classification is of the order of $80\%$.
For fainter galaxies the continuum is obviously too weak and the signal too noisy 
to allow a reliable continuum fit. 
For a number of cases, the spectra showed a spatial offset between line emission 
and continuum object of up to 1\arcsec\ (Maier et al. 2003). Considering 
that the slit generally centered on the continuum source is only 1.0\arcsec\ 
wide, it seems possible that in those cases where an emission line detected 
in the FP scan could not be verified in the slit spectra (circles in Fig.\,2), 
the slit actually missed the line emitting region. In those cases the photometric 
data derived from the imaging survey would tend to yield more reliable results 
than slit spectroscopy. A more careful analysis of the offsets between the 
continuum and line emission centroids resulting in excessive slit losses is 
currently in process.

\section{Results}        % 3.

\subsection{Final object selection}  % 3.1.

The emission line galaxy candidate samples, preselected according to the criteria 
described above, are now separated into the specific redshift ranges. 
The candidates are then subjected to a further selection procedure, where 
objects classified as quasars or stars in the 
multi-color classification are rejected. Both object types are also suspicious 
by very low overall probabilities $P$ of the line classification.

\begin{figure}[ht!]               % Fig 3
\centerline{
\psfig{figure=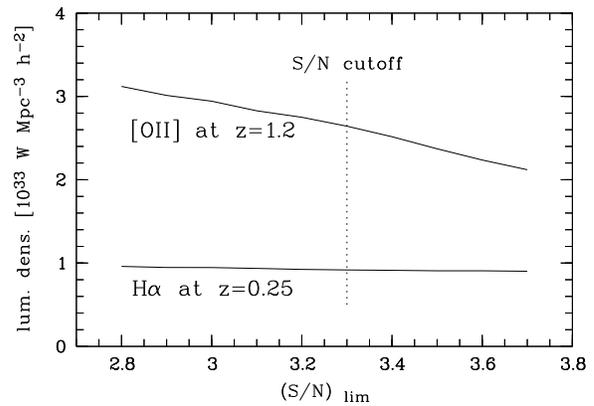,clip=t,width=7.8cm,angle=0}}
\caption[]{Line luminosity density plotted versus the chosen signal to noise limit 
for the [O\,{\sc ii}] galaxies at $z=1.2$, and for the H$\alpha$ galaxies at $z=0.25$. 
Data are extinction corrected (see Sect. 3.5), but not completeness corrected.}
\end{figure}

Fig.\,3 shows that the dependence of the resulting line luminosity density on the 
choice of the S/N limit for the emission line is not severe. 
As mentioned, a value of 3.3 seems to be appropriate. 
Lowering the cutoff line to $S/N=3.0$, which would 
include many artifacts and wrong identifications (see Fig.\,2), 
the luminosity density in the case of the $z=1.2$ sample increases by only 
about 10\%. For the galaxy sample at redshift 
0.25, which is more complete at the faint end, the effect is even smaller. 
The choice of the cutoff mainly affects the completeness at the low luminosity 
end of the line luminosity function.

\subsection{Sample statistics}          %  3.2

The total numbers of objects in the fields studied, classified as H$\alpha$ 
galaxies, as [O\,{\sc iii}]$\lambda$500.7 galaxies, and as [O\,{\sc ii}] galaxies at the 
appropriate redshifts is listed in Table 3. 

The number for galaxies seen in the other emission lines is too small for a 
useful statistical study. The expected number of galaxies detected by their 
[O\,{\sc iii}]\,495.9 line in the B window, for example, can be readily estimated from 
the statistics of the brighter [O\,{\sc iii}]\,500.7 line. Since the line intensity 
ratio of these lines is 3:1, the expected number is about that of the number 
of [O\,{\sc iii}]\,500.7\,nm with a S/N greater than $3\times(S/N)_{\rm lim}=10.5$, which 
is only 7 galaxies. Therefore, in the following, [O\,{\sc iii}] always stands for 
the brighter, the [O\,{\sc iii}]\,500.7 line. 
A similar argument leads to a total of 8 for the expected 
number of H$\beta$ line galaxies.

\begin{table}[h]                            % Tab 3
\caption{Statistics for the emission line galaxies observed in the five 
redshift intervals (from 3 CADIS fields each).}
\begin{center}
\begin{tabular}
{     c   |    c   |        c          |      c      |   c   |   c    }
 \hline   
 \hline
          &        &                   &             &       &        \\[-3mm] 
   Line   & Window &     Total         &   z range   &  S/N  &   N    \\
          &        &      area         &             & limit &        \\
          &        &                   &             &       &        \\[-3mm]
 \hline
          &        &                   &             &       &        \\[-3mm] 
H$\alpha$ &    B   & 309\,$\sq\arcmin$ & 0.238-0.252 &  3.3  &  92    \\
%H$\alpha$ &   C   & 300\,$\sq\arcmin$ & 0.390-0.411 &  3.3  &  40    \\
~ [OIII]  &    A   & 300\,$\sq\arcmin$ & 0.390-0.411 &  3.3  &  44    \\
~ [OIII]  &    B   & 309\,$\sq\arcmin$ & 0.626-0.646 &  3.3  &  80    \\
~ [OII]   &    A   & 304\,$\sq\arcmin$ & 0.867-0.894 &  3.3  & 103    \\
~ [OII]   &    B   & 309\,$\sq\arcmin$ & 1.175-1.210 &  3.3  & 119    \\[-3mm]
          &        &                   &             &       &        \\
 \hline
\end{tabular}
\end{center}
\end{table}

The range of the equivalent widths at rest wavelength in which the emission 
line galaxies are detected, is plotted vs. M$_{\rm B}$ in Fig. 4 for the H$\alpha$ 
galaxies at redshift 0.25 and for the [O\,{\sc ii}] galaxies at $z=1.2$. 
The absolute blue magnitude is calculated by interpolating between the flux 
densities from the medium and broad band filters left and right of $440(1+z)$\,nm, 
and calibrating with Vega. 
The dotted lines indicate the sensitivity limit of the survey, at 
$F_{\rm line} = 3\times 10^{-20}$ W\,m$^{-2}$. Note that our minimum detection 
limit $EW>0.2$\,nm caused by photometric uncertainties, does not affect the 
sample statistics. 

The distribution shown in this figure is consistent with the EW distribution of 
the CFRS sample (Hammer et al. 1997, Tresse \& Maddox 1998). The 
mean line equivalent width of the Tresse \& Maddox sample at $z\sim0.2$ is 
$<$log $EW$(H$\alpha$+[NII])$>$=1.59\,nm as compared to our mean value of 1.54. 
The considerably higher equivalent widths seen in nearby galaxy samples (Gallego 
et al. 1996, Ho et al. 1997, Terlevich et al. 1991) are not comparable, since their 
spectrograph slits cover only the central parts of these galaxies.

\begin{figure}[ht!]                    % Fig 4
\centerline{\vbox{
\psfig{figure=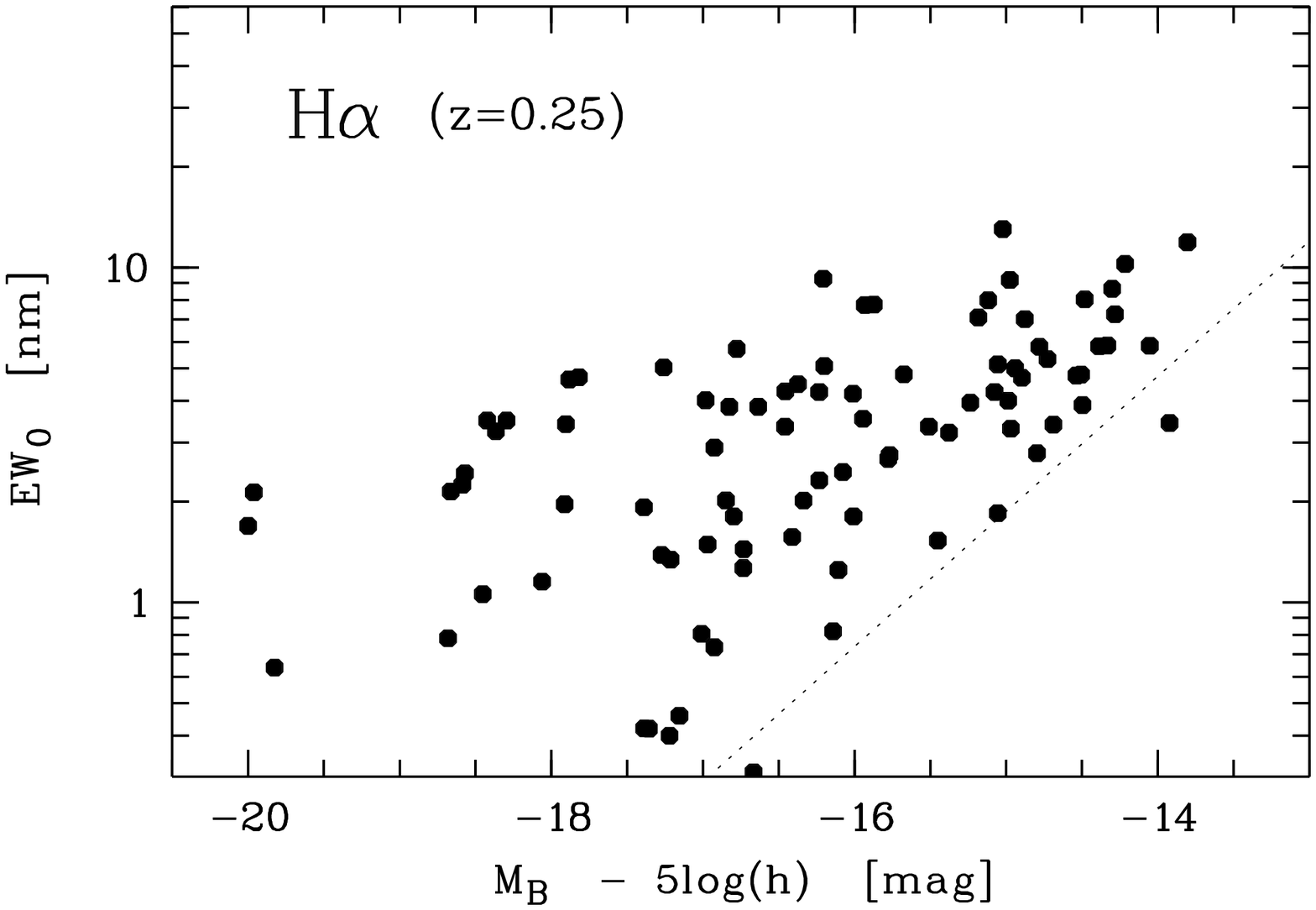,clip=t,width=8.0cm,angle=0}\vspace*{1mm}
\psfig{figure=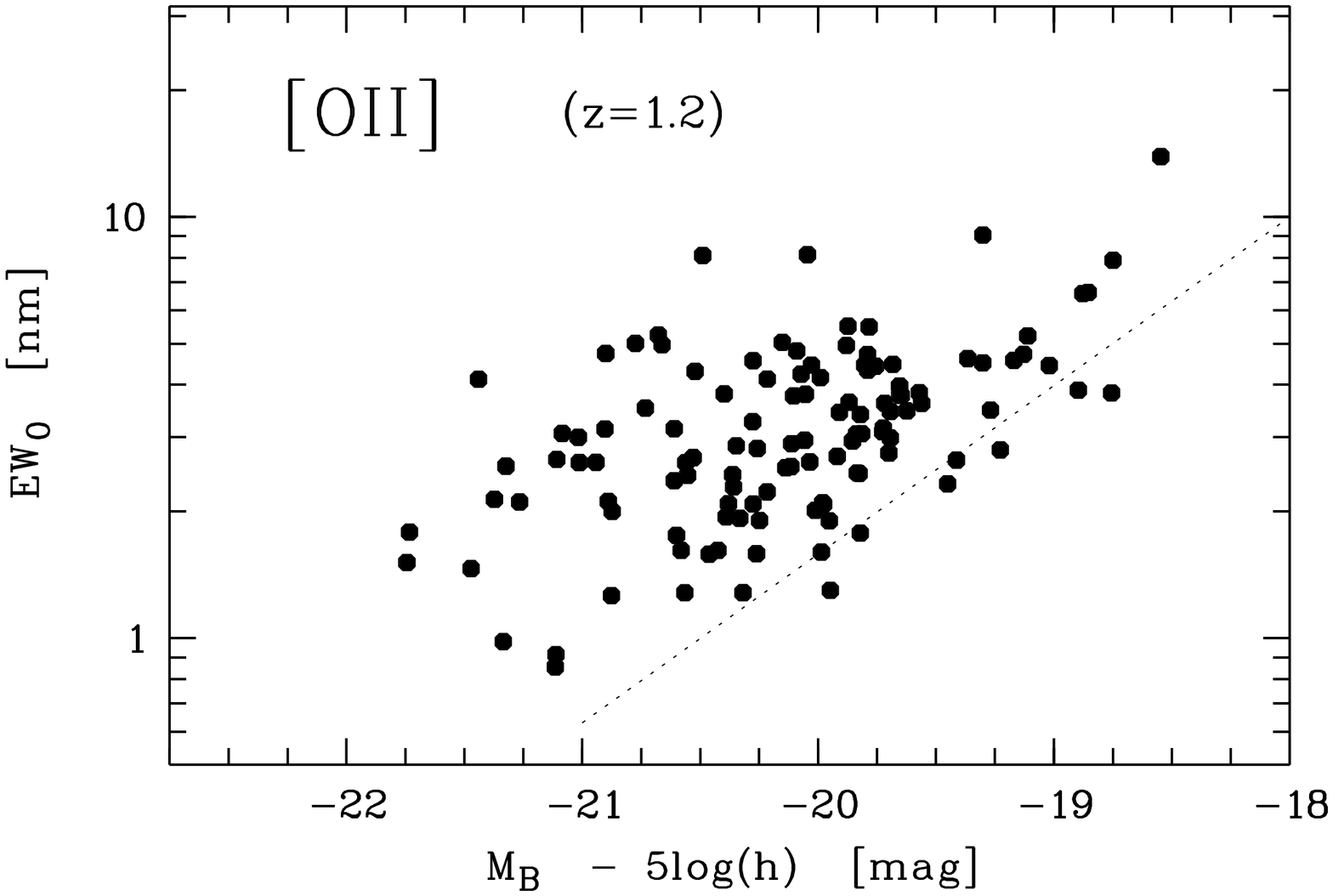,clip=t,width=8.0cm,angle=0}}}
\caption[]{H$\alpha$ and [O\,{\sc ii}]372.7 line equivalent widths at rest wavelength 
versus absolute blue magnitude; the dashed lines indicate the sensitivity limit 
of the emission line survey, at $F_{\rm line} = 3\times 10^{-20}$ W\,m$^{-2}$.}
\end{figure}

For local galaxies Kennicutt (1992) established a mean ratio between [O\,{\sc ii}] and 
H$\alpha$ eqivalent widths of 0.5. 
For galaxies identified in the B window by their H$\alpha$ line ($z=0.25$), we find a 
considerably higher ratio of ~$\sim 0.9$. 
This finding agrees with the CFRS results by Hammer et al. (1997), who explain 
the difference by a lower metallicity and hotter star temperatures in their 
$z\sim0.3$ sample, and/or by the lower luminosities as compared to the local galaxy 
sample of Kennicutt. The latter explanation would be consistent with the result of 
Gallagher et al. (1989), who find a higher EW[O\,{\sc ii}]/EW(H$\alpha$) ratio for 
blue local galaxies as well.

\subsection{Sample contamination}             % 3.3

The contamination or incompleteness of our samples due to wrongly classified 
galaxies is of the order of $\pm$10\% (see Fig.\,2). 
One has, however, also to consider contamination by galaxy types not included 
in our ELG templates, such as galaxies seen in [O\,{\sc i}] or in Mg\,{\sc ii} in 
the FP spectra, and AGNs in general. 

For discriminating these contaminants, we selected those emission line galaxies for 
which the CADIS multi-color classification (Wolf et al. 2001) yields a different 
line solution or indicates a quasar. After eye inspection of the spectra, 
two galaxies initally classified at $z=0.25$ were found to 
be at redshift 0.30 ([O\,{\sc i}] at $\sim$820nm). Two possible quasars at redshift 
1.2 and another two at 1.9 (Mg\,{\sc ii} in window B) were recognized and removed 
from the [O\,{\sc ii}] sample. Among the ELGs seen in the A window no QSO was found. 

The line ratios measured by the CADIS emission line survey have relatively large 
error bars and do not allow to recognize Seyfert 2 galaxies, neither does the multi-color 
classification. 
Spectroscopy of AGNs in the Chandra deep fields indicate a surface density of 
450 Sy2 galaxies per square degree at $R<24$ and $0.5<z<1$ (Szokoly et al., 2002), 
equivalent to only three Sy2 in our $z=1.2$ sample. At lower redshifts even fewer 
objects are expected. 

LINERS are, among other, recognizeable by a high [N{\sc ii}]/H$\alpha$ line flux 
ratio. The line fits for the 30 strongest line emitters in our $z=0.25$ sample 
yield two LINERS ($EW$[NII]/$EW$(H$\alpha$)$>$0.6). The contamination of the total 
H$\alpha$ luminosity by the [NII] fluxes from these two LINERs is about 2\%. 
In the CFRS no low-excitation AGN was found (Hammer et al. 1997). Thus, we are 
confident that the LINER contamination is negligible in all our samples.

\subsection{Luminosity functions}              %  3.4

Fig.\,5 shows the $M_{\rm B}$ luminosity functions for the H$\alpha$ and [O\,{\sc ii}] galaxies 
observed in window B (average over 3 CADIS fields), and for the [O\,{\sc ii}] galaxies 
observed in window A (average over 3 fields). 
The data points for the mean density are in agreement to the luminosity function 
obtained for the late type galaxies (later than Sa) in the respective 
redshift ranges, derived from the multi-color survey of CADIS (Fried et 
al. 2001). 
The Schechter functions overlaid in the figure are also adopted from Fried et al. 
and represent the sum of the two LFs for starburst galaxies and spirals in 
their paper. 

This agreement suggests that in principle all late type galaxies at these 
redshifts are emission line galaxies with lines bright enough to be detected 
in the CADIS Fabry-Perot observations.

\begin{figure}[ht!]               %5
\centerline{\vbox{
\psfig{figure=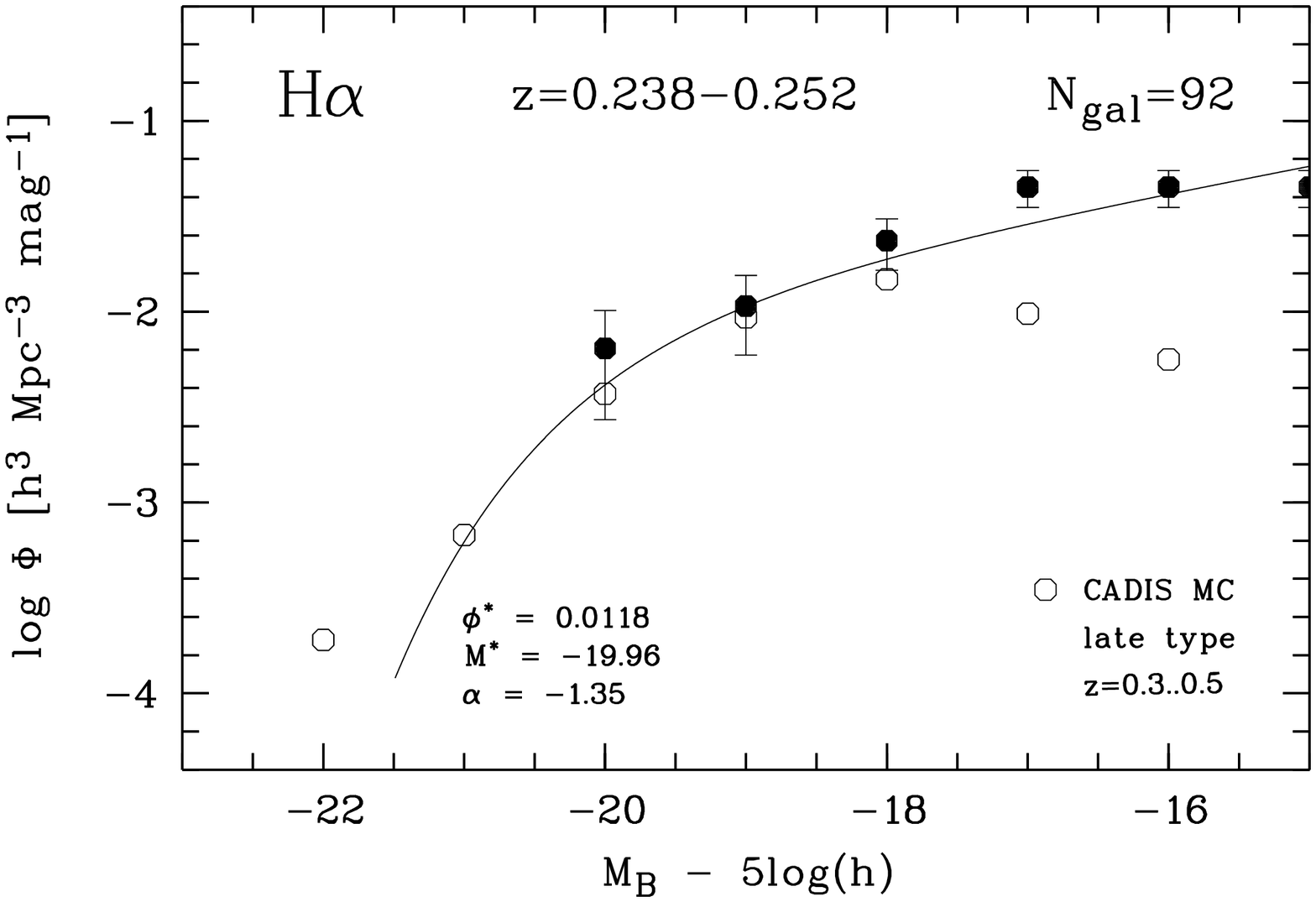,clip=t,width=8.0cm,angle=0}\vspace*{1mm}
\psfig{figure=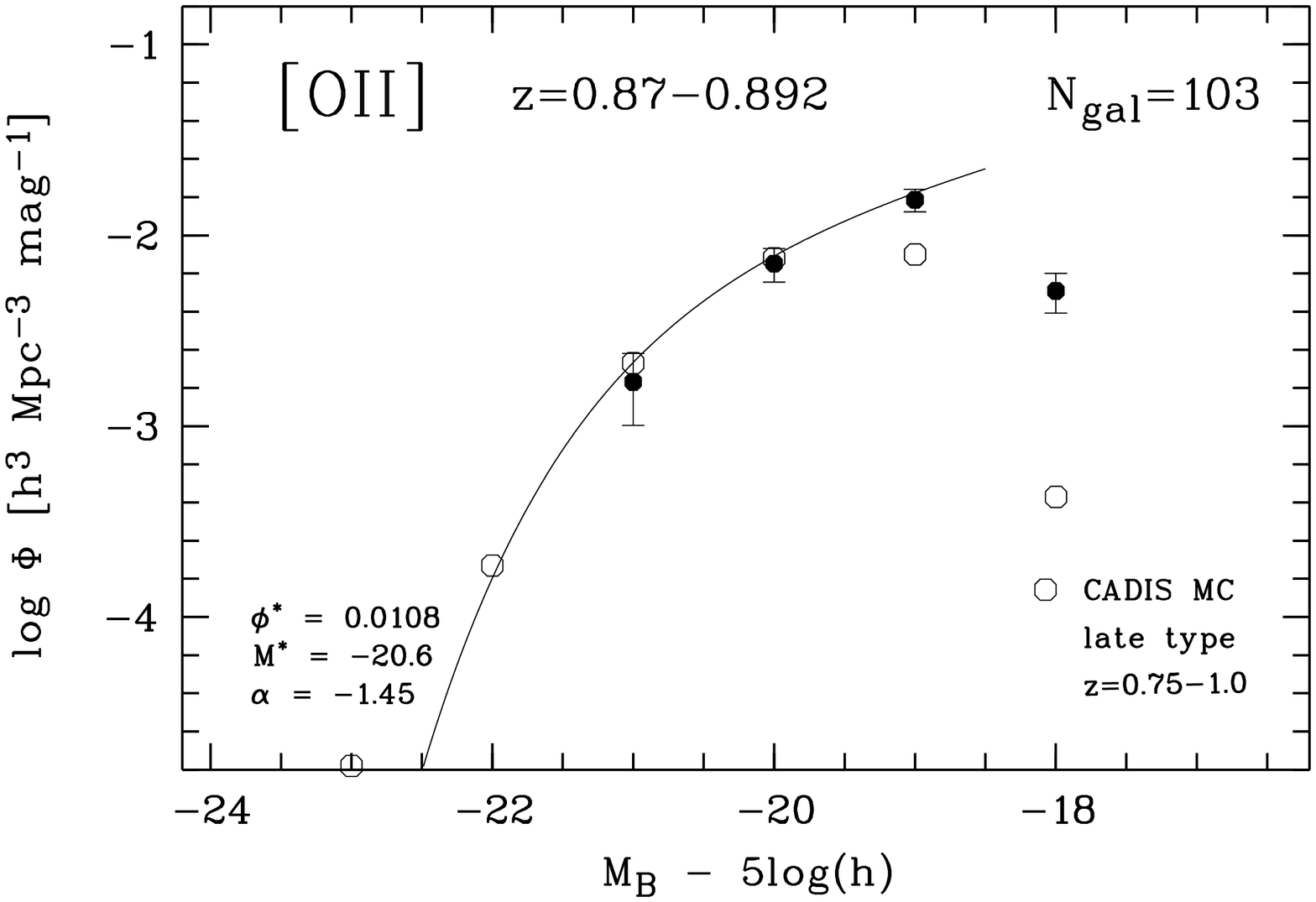,clip=t,width=8.0cm,angle=0}\vspace*{1mm}
\psfig{figure=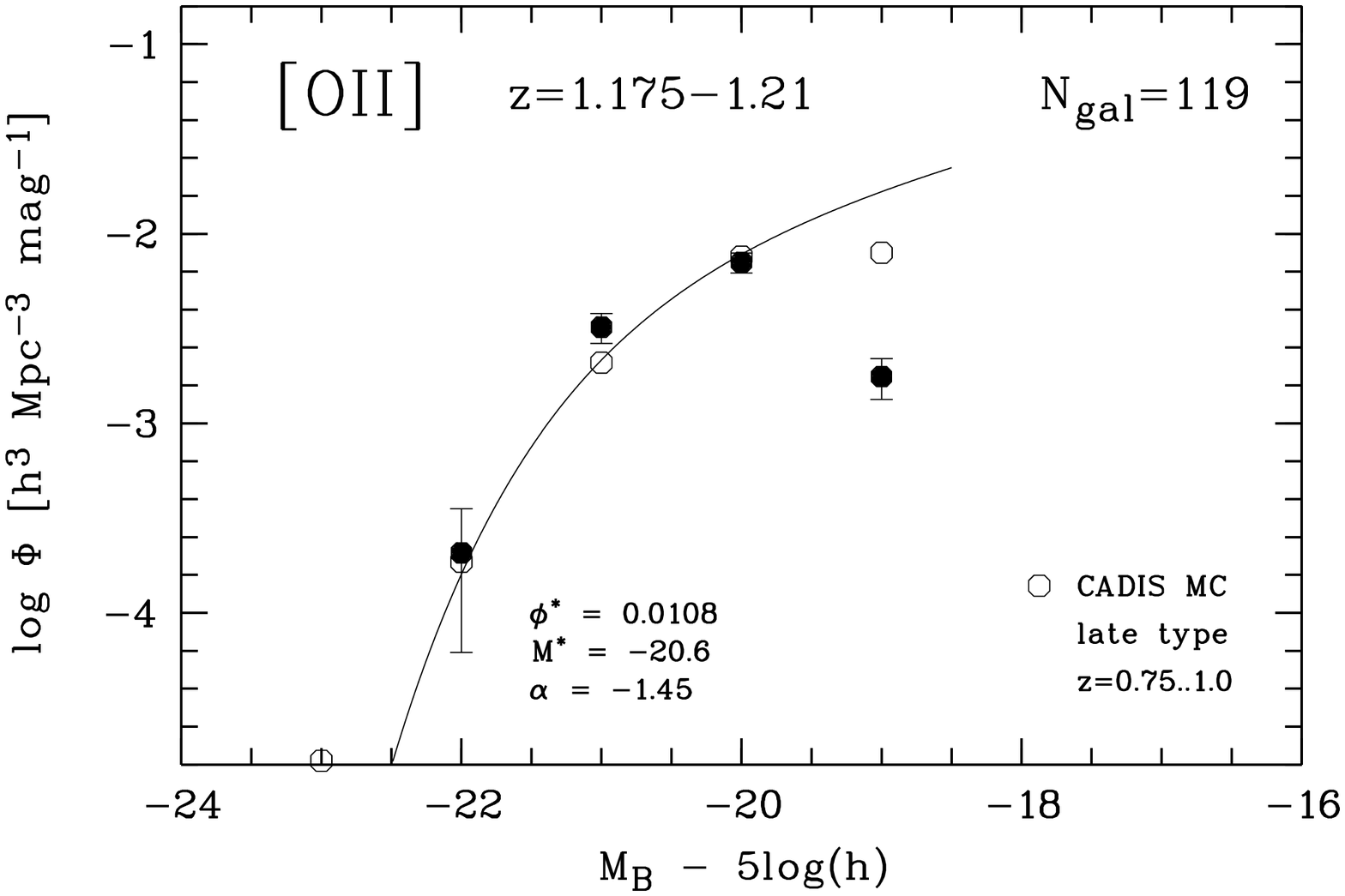,clip=t,width=8.0cm,angle=0}}}
\caption[]{Luminosity functions for the emission line galaxies at $z=0.25$, 
0.64, and 1.2 (filled circles). For comparison open circles trace the luminosity 
functions for late type galaxies according to the CADIS multi-color survey (Fried 
et al. 2001), together with the corresponding Schechter functions. Schechter 
parameters are given at the high luminosity ends of the curves. } 
\end{figure}

\subsection{Extinction correction}       %  3.5.

Since star forming regions are generally embedded in massive and dusty 
molecular clouds, extinction plays an important role in the determination 
of intrinsic line emission and can amount to more than a magnitude in 
the H$\alpha$ line. 

The calibration factors used to convert the line luminosity densities 
into star formation rates are derived for unobscured line emission (Kennicutt 
1983, 1992). So our line luminosities need to be corrected for extinction 
before deriving star formation rates. This correction is generally based on the 
H$\beta$/H$\alpha$ intensity ratio. In our data sample, the fluxes for 
the H$\beta$ line contain large relative errors, since the flux contained 
in the H$\beta$ line is in most cases not directly measurable, but only 
as blend with the [O\,{\sc iii}] lines. Thus, a reddening correction for each 
individual galaxy is not possible. 

There is, however, an observed trend in the sense that more luminous galaxies 
(disks) show higher reddening than less luminous ones, such as H\,{\sc ii} 
galaxies, irregulars, BCDs (Gallego et al. 1997), which allows to perform a 
{\it global} correction based on the blue magnitude of each galaxy. 
This trend is clearly seen in the nearby galaxy sample used by Calzetti et al. 
(1994) to study the internal reddening. Their data suggest an increase of 
reddening according to 
\begin{center}
$E_{\rm B-V} = 0.10 ~ (M_{\rm B}+15.5) ~ \pm 0.15$. 
\end{center}
with no difference between spirals and BCGs. 

For a larger galaxy sample of nearby galaxies, Jansen et al. (2001) investigated the 
dependance of the [O\,{\sc ii}]/H$\alpha$ emission-line ratio on other observables. 
They found that the large scatter of the observed ratio and its dependence 
on luminosity are greatly reduced when using reddening-corrected [O\,{\sc ii}] fluxes. 
Their correlation between absolute magnitude and reddening, derived from the Balmer 
decrement after correction of the underlying stellar Balmer absorption, indicates 
a relation with a less steep slope: 

\begin{center}
  $E_{\rm B-V} = 0.064 ~ (M_{\rm B}+14.8) ~ \pm 0.10$. 
\end{center}
independent of the galaxy type. 
The large uncertainty, which is engendered by the large scatter of the data points 
and due to geometrical and excitation temperature effects, makes it impossible to 
predict the reddening for a particular galaxy. 
Since the present study, however, deals with global parameters instead of 
individual galaxies, it is appropriate to apply this relation for 
an averaged reddening correction. 
For faint galaxies ($M_{\rm B}>15.6$), we adopt a minimum reddening correction 
of $E_{\rm B-V}=0.05$\,mag.

\subsection{Luminosity functions for the emission line fluxes}     %  3.6

The next step towards establishing global star forming rates is the determination 
of luminosity functions for the emission line fluxes. Fig.\,6 shows these for 
the same three redshift intervals as in Fig.\,5. 
In these diagrams, the sampling interval for the line luminosity was chosen as 
0.4\,dex, equivalent to 1 magnitude stepsize, in order to allow an easy 
comparison with the $M_{\rm B}$ luminosity functions. 

In the $L_{\rm H\alpha}$ plot (top) the observed line luminosity data of Tresse 
\& Maddox (1998) and by Cowie et al. (1997), after being converted to the 
cosmological parameters used in the present paper, are over plotted. In the 
$z=1.2$ panel, the Cowie et al. (1997) points are over plotted. Both agree 
satifactorily well with our data.

The extinction corrected luminosity functions are shown by filled symbols. 
For the case of the H$\alpha$ luminosity at $z=0.25$, the correction yields an 
increase of the total line luminosity by a factor of 1.6. For the $z=1.2$ sample 
the extinction correction amounts to a factor of about 4 in the total [O\,{\sc ii}] 
line luminosity (see Table 4), due to both the higher blue luminosities of the 
galaxies and the shorter wavelength of the line.

\begin{figure}[h!]                   % Fig 6
\centerline{\vbox{
\psfig{figure=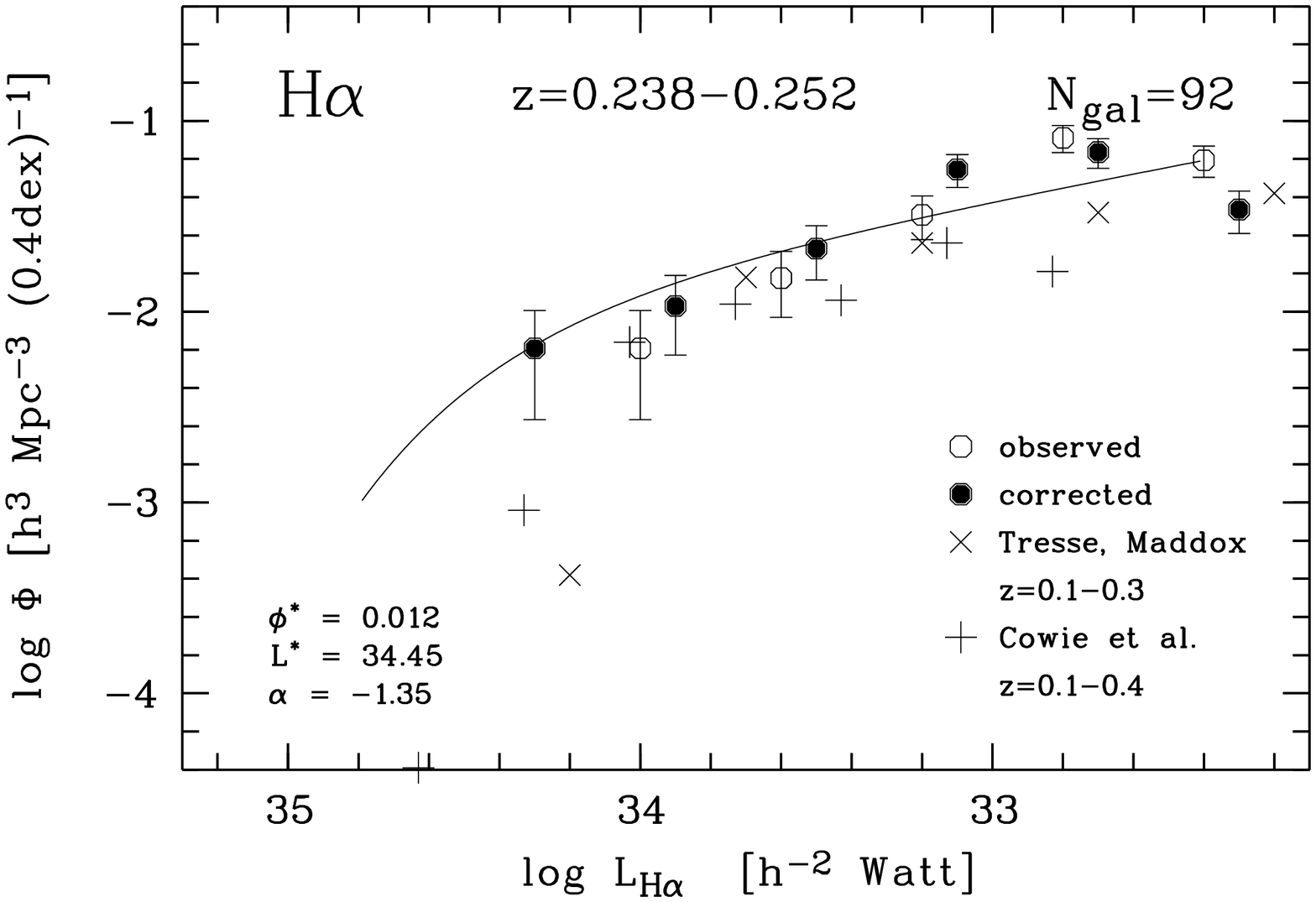,clip=t,width=8.0cm,angle=0}
\psfig{figure=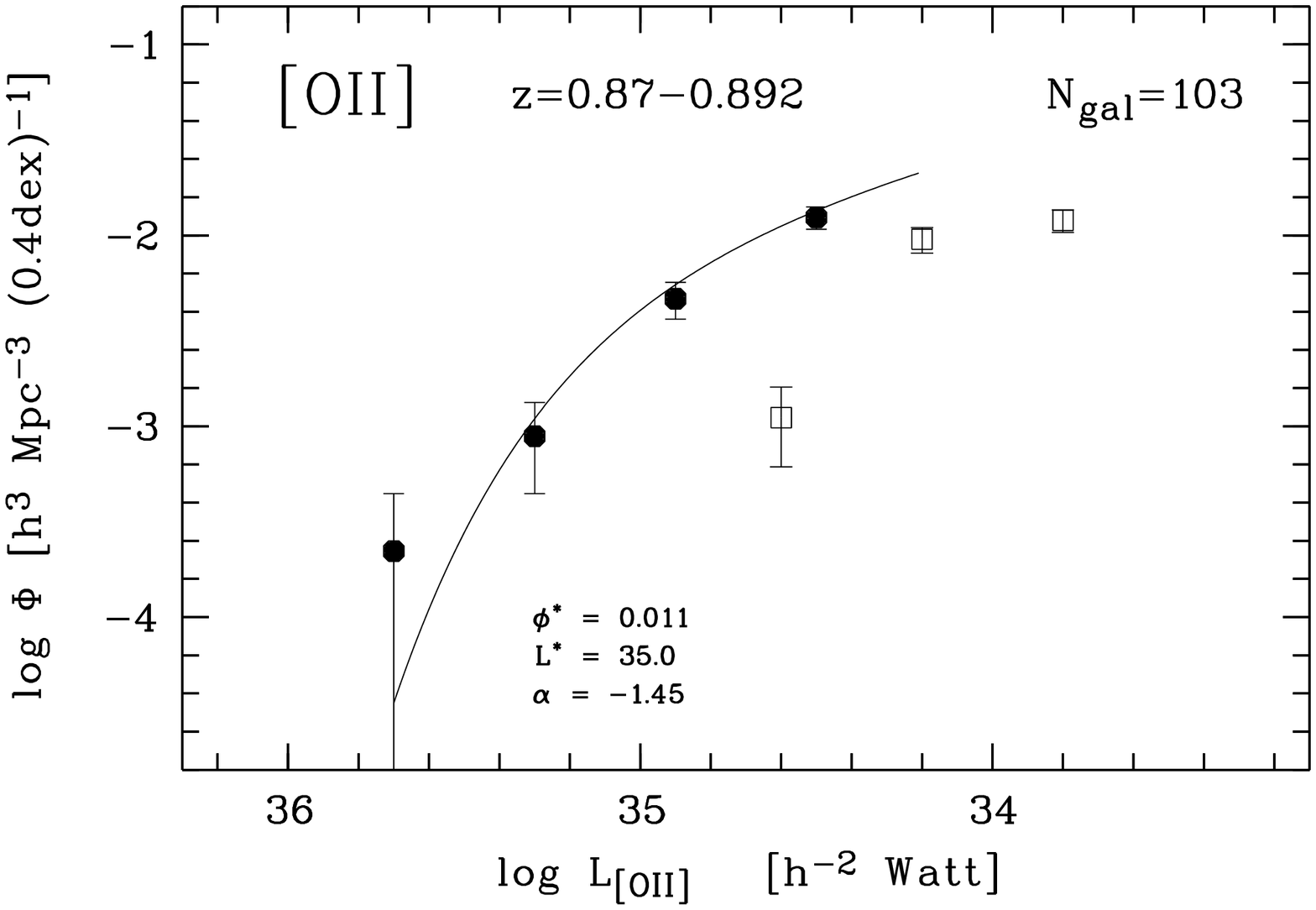,clip=t,width=8.0cm,angle=0}
\psfig{figure=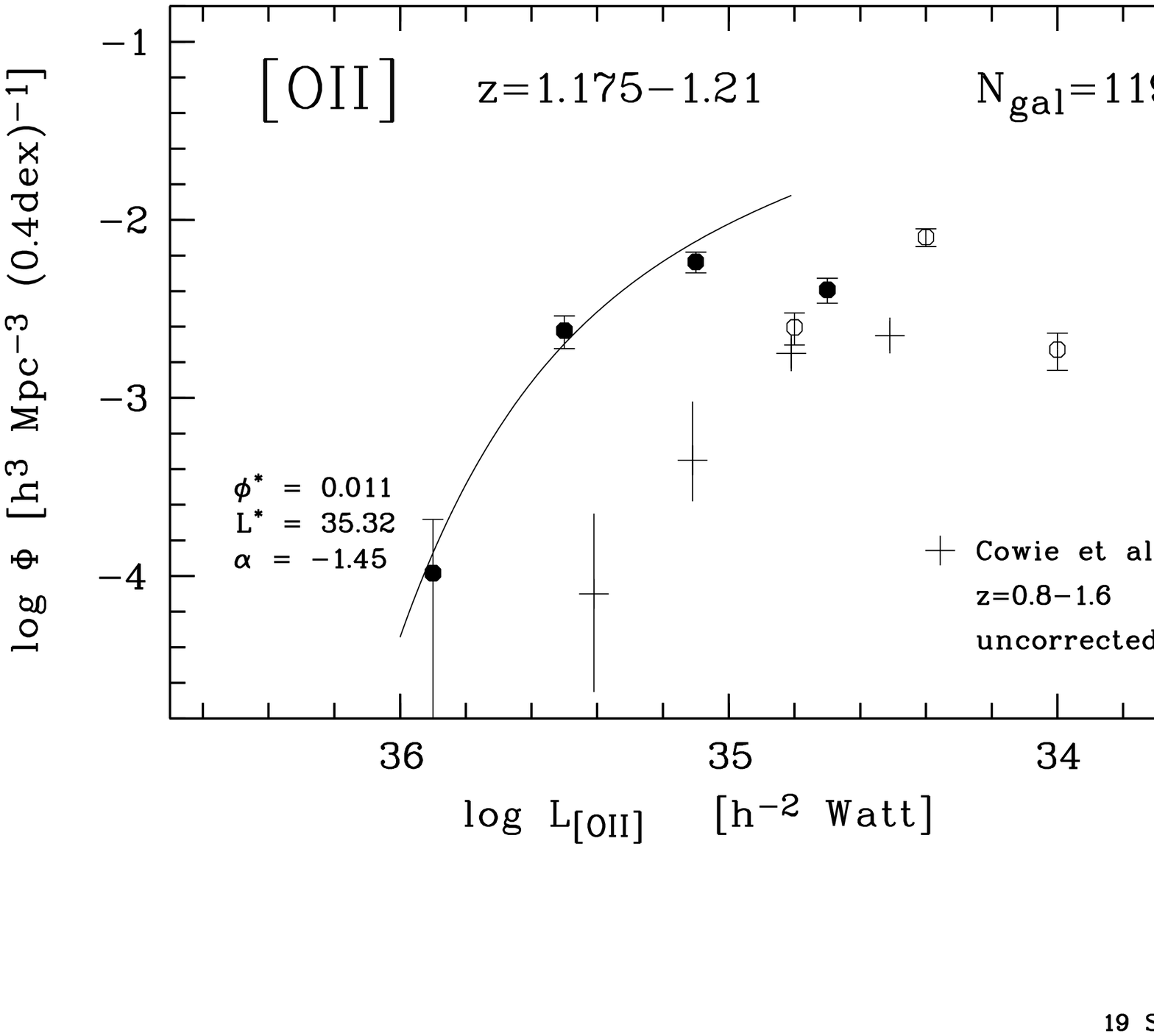,clip=t,width=8.0cm,angle=0}}}
\caption[]{Luminosity functions for the observed H$\alpha$ line emission 
at redshift 0.25 (top), for the [O\,{\sc ii}] line at redshift $z = 0.88$, and 
for the [O\,{\sc ii}] line at redshift $z = 1.2$; open cirles are uncorrected, filled 
circles are extinction corrected values; for comparison, the observed (uncorrected) 
line luminosity function by Cowie et al. (1997) and by Tresse \& Maddox (1998) 
are shown. The parameters for the Schechter 
functions plotted over the data have the same $\Phi$ and $\alpha$ as the 
M$_{\rm B}$ luminosity functions. }
\end{figure}

In several studies (Gallego et al. 1995, Tresse \& Maddox 1998, and Yan et al. 
1999) line luminosity functions were successfully described by Schechter functions, 
the parameters of which were derived by the V$_{\rm max}$ method. 
For small data samples, as it is the case in the present study, this procedure 
leads however to arbitrary results. 

As can be seen in the plots, the H$\alpha$ luminosity density histogram (z=0.25) 
can be described with a Schechter function of the same shape 
as the respective blue magnitude function. The over plotted curves in Fig. 5 
are fits to the observed distribution using the same $\Phi^{\ast}$ and $\alpha$ 
parameters as in the corresponding $M_{\rm B}$ luminosity functions.  
The reason for this agreement seems to be, that on the one hand the fainter galaxies are 
preferentially of type BCDs or Irr, leading to a steepening of the line luminosity 
function as compared to the blue magnitude function. On the other hand, the 
internal extinction acts preferentially on massive galaxies (Sect. 3.5), 
and thus its correction yields a stretching the luminosity function at its bright 
end, and thus partly cancelling the other effect. 
Therefore, we use for the Schechter function of the line luminosity functions 
the same shape parameters as those describing the blue magnitude 
luminosity functions. 

We will first discuss the luminosity functions for the H$\alpha$ and for 
the [O\,{\sc ii}] lines, which can more or less directly be used for the determination 
of the star forming rate. The [O\,{\sc iii}]\,501 line is discussed in Sect. 4.2.

\section{Discussion}   % 4.

\subsection{Star forming rate from emission lines}  % 4.1.

As shown by Kennicutt (1983) the total number of ionizing photons 
by newly produced stars is a good measure for the current star 
formation in a galaxy. Using updated evolutionary track, and 
assuming case B recombination and for a Salpeter IMF, 
Kennicutt (1998) derived a conversion factor for integral H$\alpha$ 
luminosity to star formation rate of 
\begin{center}
$ {\rm SFR} = 0.79 \times 10^{-34} L_{\rm H\alpha}$ [M$_{\odot}$\,yr$^{-1}$\,W$^{-1}$]. 
\end{center}

A relation with similar conversion factor can be used for the galaxies 
seen in the light of the [O\,{\sc ii}] line. For nearby galaxies, Kennicutt (1992) 
derived, for extinction corrected [O\,{\sc ii}] luminosities 
${\rm SFR} = 5 \times 10^{-34} L_{\rm [O\,II]}$ [M$_{\odot}$\,yr$^{-1}$\,W$^{-1}$]. 
Gallagher et al. (1989) for blue galaxies and Cowie et al. (1997) 
for rapidly star-forming galaxies found a considerably lower conversion 
factor of  $1.0 \times 10^{-34} L_{\rm [O\,II]}$\,M$_{\odot}$\,yr$^{-1}$/W. 
For high redshifdt galaxies, Thompson \& Djorgovski (1991) used a factor 
of $1.2 \times 10^{-34}$ for the [O\,{\sc ii}] line emission.

\begin{figure}[ht!]                  % Fig 7
\centerline{
\psfig{figure=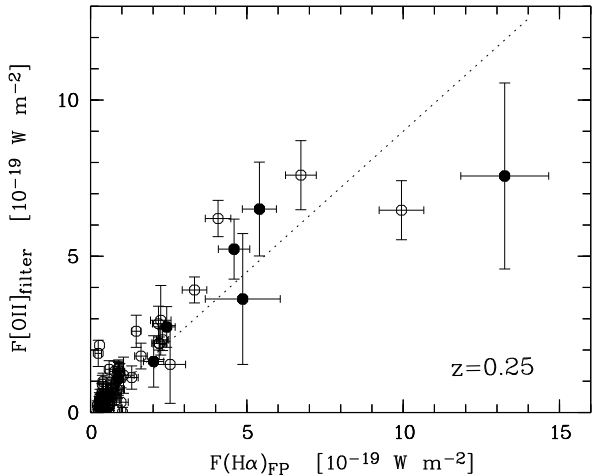,clip=t,width=8cm,angle=0}}
\caption[]{Extinction corrected [O\,{\sc ii}] line fluxes plotted over the 
H$\alpha$ line fluxes for the galaxy sample at $z=0.25$. The mean ratio 
between both line fluxes is displayed by a straight line with 
$F${\rm [O\,{\sc ii}]}/$F{\rm (H\alpha)}=0.9$. Filled circles indicate 
galaxies with $M_{\rm B} < -18.5$. 
}
\end{figure}

For galaxies seen in the light of H$\alpha$ at $z=0.25$, the [O\,{\sc ii}] line 
fluxes are measured with the CADIS filter at $\lambda_{\rm c}=465$\,nm. 
(see top of Table 1). 
In Fig.\,7 the line fluxes are plotted for the galaxies with 
$(S/N)_{\rm [O\,II]}>1$. The average flux ratio results in 
$F{\rm [O\,II]}/F{\rm (H\alpha)} \sim 0.9$, in good agreement 
with the values cited above. 
This ratio is roughly in the mean of the ratios derived by Kennicutt (1998) 
and Sullivan et al. (2000).

Although there is a difference of roughly 4\,Gyr between the epochs 
for $z=1.20$ and $z=0.25$, the line ratio may not have changed 
significantly. Based on the CFRS galaxy sample with redshifts in the range 
$0.5<z<1.0$, Carollo \& Lilly (2001) found a remarkable similarity of 
metallicities to that for local galaxies. Also, Pettini et al. (1998) 
and Teplitz et al. (2000) found the line spectra at high redshift to 
be very similar to those observed in local emission line galaxies. 
The calibration used in the present study for the [O\,{\sc ii}] emission is  
therefore
\begin{center}
$ {\rm SFR} = 0.88 \times 10^{-34} L_{\rm [O\,II]}$ ~ [M$_{\odot}$\,yr$^{-1}$\,W$^{-1}$]. 
\end{center}

Rosa-Gonzalez et al. (2002) derived SFR estimators using the Calzetti 
extinction curve and accounting for the underlying stellar Balmer absorption. 
Applying their calibration to our non-extinction corrected [O\,{\sc ii}] 
galaxy sample we would end up with about two times higher star forming rates.

\subsection{Star forming rates from the [O\,{\sc iii}]\,$\lambda$500.7 line.}   % 4.2. 

Having detected and measured a large number of [O\,{\sc iii}] emission lines 
at medium redshift, we will test how far this prominent line can be used 
as SFR estimator. 

The line luminosity functions for the galaxies observed in the [O\,{\sc iii}] line 
at $z=0.40$ and $z=0.64$ are shown in Fig.\,8. Again the distributions are 
fitted with Schechter functions, the parameters $\phi$ and $\alpha$ of which are 
taken from the $M_{\rm B}$ luminosity functions for the same redshift range (Fried et 
al. 2001).

\begin{figure}[ht!]                   % Fig 8
\centerline{\vbox{
\psfig{figure=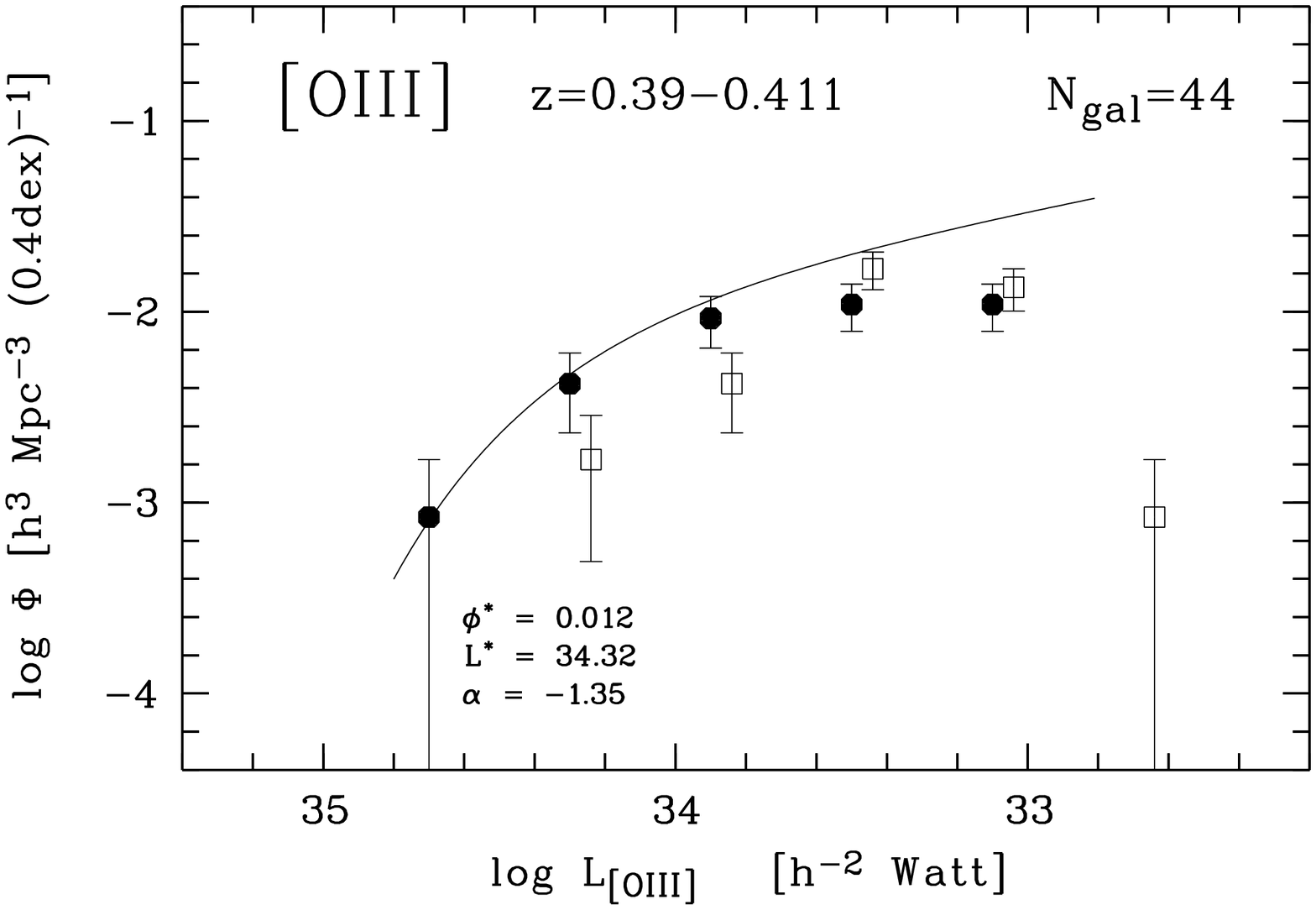,clip=t,width=8.0cm,angle=0}
\psfig{figure=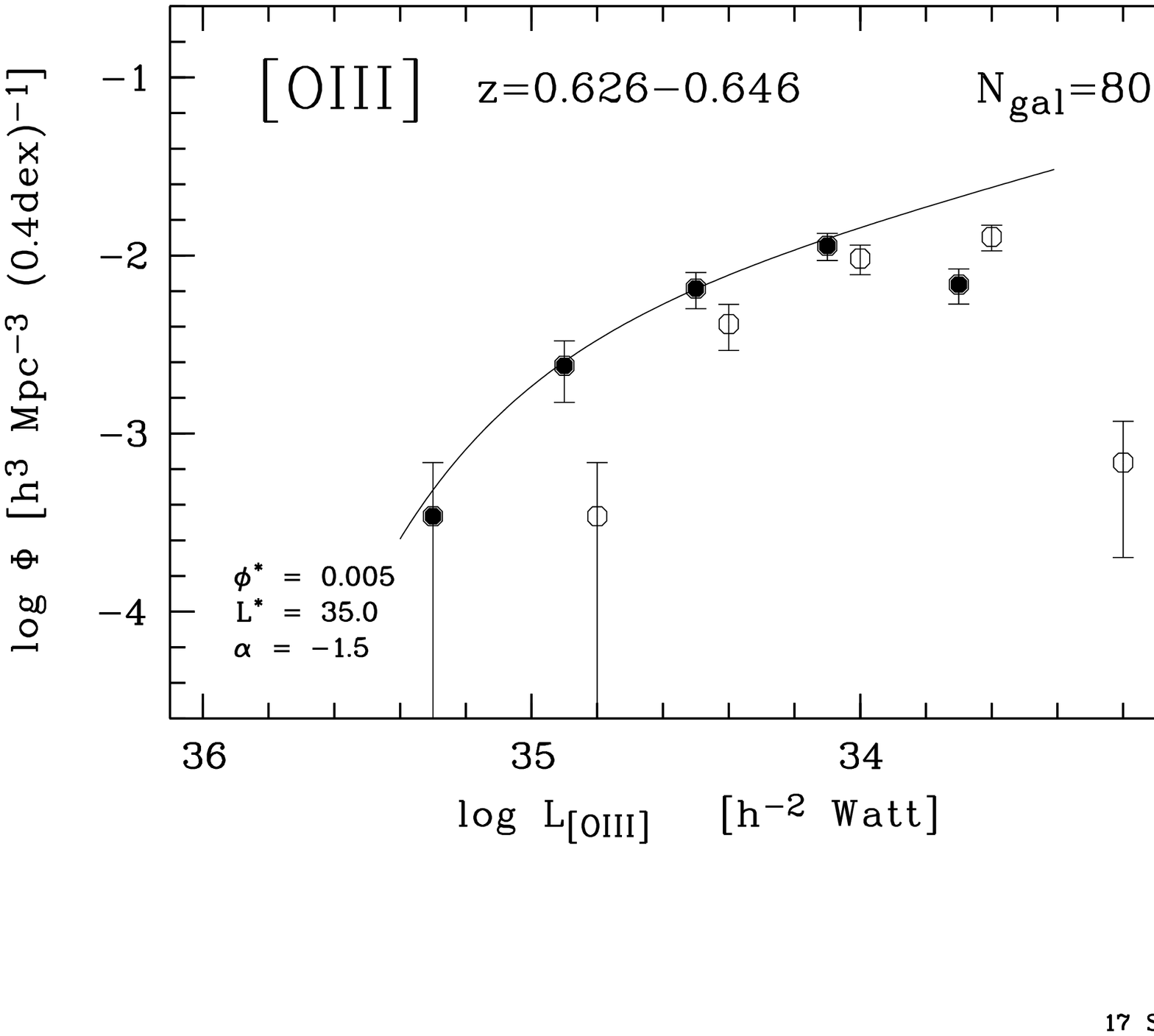,clip=t,width=8.0cm,angle=0}}}
\caption[]{Luminosity functions for the [O\,{\sc iii}] line emission 
at redshift 0.40 (top), and at redshift 0.64; open cirles are uncorrected, filled 
circles are extinction corrected values. Parameters for the Schechter function 
drawn are listed at the bright end of the LF. 
}
\end{figure}

Due to its high ionization level the luminosity of [O\,{\sc iii}]\,500.7 depends 
strongly on excitation and metallicity. 
In order to convert the [O\,{\sc iii}] emission line luminosities into star formation 
rates, an averaged intensity ratio between H$\alpha$ and the ogygen line 
has therefore to be established. Since the population of galaxies may change with 
redshift, one has to use data sets measured at comparable redshifts for this.

\begin{figure}[ht!]                  % Fig 9
\centerline{
\psfig{figure=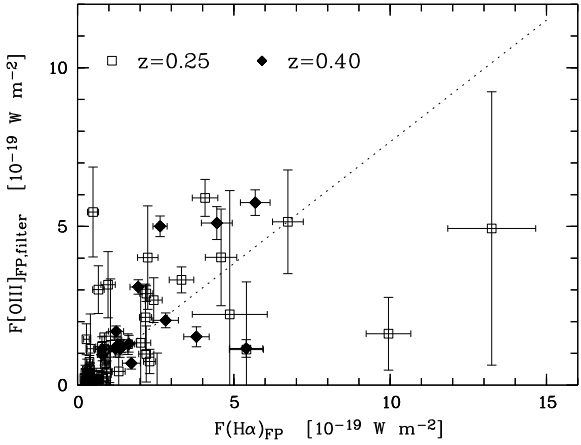,clip=t,width=8cm,angle=0}}
\caption[]{Extinction corrected [O\,{\sc iii}] line fluxes plotted over the H$\alpha$ 
line fluxes for our galaxy sample at $z=0.25$ and at $z=0.40$; filled symbols 
are for galaxies with $M_{\rm B}<-18.5$\,mag. The dotted line indicates the mean 
line ratio. 
}
\end{figure}

For the galaxies observed at $z=0.25$ in H$\alpha$, the extinction corrected 
fluxes for [O\,{\sc iii}]\,500.7 are plotted versus the H$\alpha$ fluxes as diamonds 
in Fig.\,9. Also shown in this graph, as squares, are the corrected line flux data for 
a sample of star forming galaxies at $z=0.40$. At this redshift, 
the H$\alpha$ line happens to occur in the atmospheric C 
window at $\lambda \sim 918$\,nm and, simultaneously, the [O\,{\sc iii}] line in the 
A window at $\lambda \sim 702$\,nm. Thus, the fluxes for both lines can be 
derived from FP scans rather accurately. 

While the data points in the [O\,{\sc ii}]/H$\alpha$ plot (Fig. 7) appear nicely 
correlated - in accordance with the well defined sequence in the diagnostic diagram 
[O\,{\sc ii}]\,372.7/[O\,{\sc iii}]\,500.7 vs. [O\,{\sc iii}]\,500.7/H$\beta$ 
(Baldwin et al. 1981) -, the data points in the H$\alpha$ vs. [O\,{\sc iii}] flux 
diagram scatter considerably. The mean flux ratio determined from the global 
line emission densities is 0.76. 

For the galaxies seen in [O\,{\sc iii}] at $z=0.64$, the [O\,{\sc ii}] line shows up 
in the medium band filter at 611\,nm (see Fig.\,1, second panel). The mean flux ratio 
derived from the global line emission densities of these galaxies is 
$F$[O\,{\sc iii}]/$F$[O\,{\sc ii}]$ \approx 0.9$, in good agreement with the 
[O\,{\sc ii}]/H$\alpha$ and [O\,{\sc iii}]/H$\alpha$ ratios derived above. 

We thus use for [O\,{\sc iii}] the calibration: 
\begin{center}
 ${\rm SFR} = 1.00 \times 10^{-34} L_{\rm [O\,III]}$ ~ [M$_{\odot}$\,yr$^{-1}$\,W$^{-1}$]. \\
\end{center}

\begin{table}
\caption{Luminosity densities and star forming rates for the five lines 
and redshift intervals studied (for $H_{\rm 0}=70$\,km\,sec$^{-1}$\,Mpc$^{-1}$). 
L is given as sum of the observed fluxes, after extincion corrected, and 
after completeness correction. 
}
\begin{center}
\begin{tabular}
%{    c         c     |  c  |   c       c       c          c          }
% \hline   
% \hline
%         &           &     & \multicolumn{3}{c}{ } &                 \\[-3mm] 
%    z    &    Line   &  N  &
% \multicolumn{3}{c}{L ~ in 10$^{32}$\,W\,Mpc$^{-3}$} &     SFR       \\
%         &           &     &  obs. & corr. & compl.&
%                                             M$_{\odot}$/Mpc$^3$/yr  \\[-3mm]
%         &           &     &       &       &       &                 \\
% \hline
%         &           &     &       &       &       &                 \\[-3mm] 
% 0.25    & H$\alpha$ &  92 & 1.65  &  2.7  &  2.9  & 0.024$\pm$0.006 \\
% 0.40    &  [OIII]   &  44 & 0.83  &  1.8  &  2.4  & 0.024$\pm$0.008 \\     
% 0.64    &  [OIII]   &  80 & 1.75  &  5.7  &  7.2  & 0.072$\pm$0.016 \\     
% 0.88    &  [OII]    & 103 & 2.00  &  7.5  & 12.2  & 0.107$\pm$0.035 \\     
% 1.20    &  [OII]    & 119 & 2.60  & 11.8  & 25.9  & 0.228$\pm$0.055 \\[-3mm]
%         &           &     &       &       &       &                 \\
% \hline
{    c         c     |    c       c       c          c          }
 \hline   
 \hline
         &           &  \multicolumn{3}{c}{ } &                 \\[-3mm] 
    z    &    Line   &  
 \multicolumn{3}{c}{L ~ in 10$^{32}$\,W\,Mpc$^{-3}$} &     SFR       \\
         &           &   obs. & corr. & compl.&
                                        M$_{\odot}$/Mpc$^3$/yr  \\[-3mm]
         &           &        &       &       &                 \\
 \hline
         &           &        &       &       &                 \\[-3mm] 
 0.25    & H$\alpha$ &  1.65  &  2.7  &  2.9  & 0.024$\pm$0.006 \\
 0.40    &  [OIII]   &  0.83  &  1.8  &  2.4  & 0.024$\pm$0.008 \\     
 0.64    &  [OIII]   &  1.75  &  5.7  &  7.2  & 0.072$\pm$0.016 \\     
 0.88    &  [OII]    &  2.00  &  7.5  & 12.2  & 0.107$\pm$0.035 \\     
 1.20    &  [OII]    &  2.60  & 11.8  & 25.9  & 0.228$\pm$0.055 \\[-3mm]
         &           &        &       &       &                 \\
 \hline
\end{tabular}
\end{center}
\end{table}

\subsection{Completeness correction}  % 4.3.

Table 4 lists the line luminosity densities as observed by summing up the data 
after extinction correction and completeness correction, combined with 
the star forming rates for $H_{\rm 0}=70$\,km\,sec$^{-1}$\,Mpc$^{-1}$. 

The completeness correction is done by calculating the area under the Schechter 
function according to: 
\begin{center}
${\rm SFR_{tot}} = \Phi^{\ast} \times L^{\ast} \times \Gamma(2 + \alpha)$ 
\end{center}

A comparison between the SFR derived from the sum of the observed galaxy line 
emissions  and the area under the Schechter function using the parameters marked 
in the plots shows that for the case of $z=0.25$, the completeness correction is 
negligible, while for the $z=1.2$ sample it amounts to a factor $>2$. 

Small uncertainties in the slope parameters $\alpha$ may lead to large errors 
in the total luminosity and SFR. For $\alpha \sim 1.4$, which is the mean value in 
our curves, a variation of the slope parameter $\alpha$ by $\pm 0.1$ would 
alter the SFR by $\pm \sim 15\%$.

\subsection{Evolution of SFR}        % 4.4.

The SFR values (Table 4) show an increase by about a factor of 10 between 
$z=0.25$ and $z=1.2$. This is in good agreement with the slope of the SFR data 
published by Hogg et al. (1998), who used the [O\,{\sc ii}] line for all 
redshifts from 0.2 to 1.2, and with the slope derived from UV flux densities 
(Lilly et al. 1996).

\begin{figure}[ht!]                   % Fig 10
\centerline{\psfig{figure=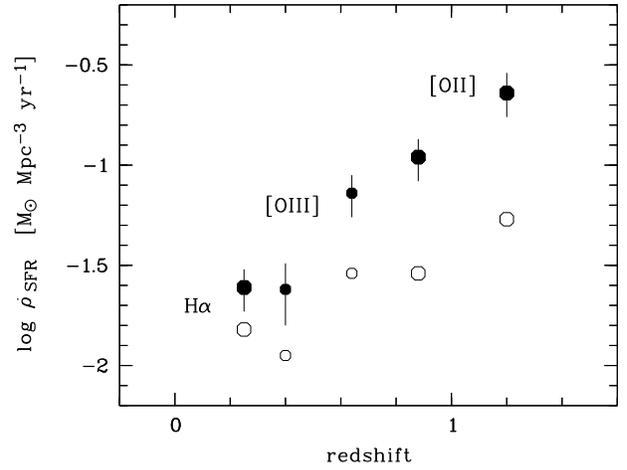,clip=t,width=8.2cm,angle=0}}
\caption[]{CADIS results for the completeness corrected star forming rate density 
as function of redshift. 
Large symbols are for H$\alpha$ ($z=0.25$) and [O\,{\sc ii}] lines ($z=0.88, ~1.2$), 
small symbols (for [O\,{\sc iii}] at $z=0.40$ and  0.64) are inserted for 
comparison. Filled circles stand for extinction-corrected, open ones for uncorrected 
values. Error bars are statistical. 
}
\end{figure}

As can be seen, the values at $z=0.88$ and 1.2 are rather sensitive to 
the extinction correction applied. The dispersion of $\pm0.1$ in the reddening 
relation (Sect. 3.5) yields a systematic error bar of $\pm40\%$ for both 
SFR values. 

The SFR densities derived from the [O\,{\sc iii}] line follow the trend of the 
other data within the errors. Using a large sample of galaxies and averaged line 
ratios, this line yields satisfactory results.

\subsection{Field to field variation}

Large scale structure 
expresses itself in strongly varying number counts from field to field and 
from redshift to redshift. This effect can nicely be seen in the $M_{\rm B} - z$ 
diagrams for the CADIS galaxies (Fried et al., 2001). Since the redshift 
intervals of our FP windows correspond roughly to the typical size of a LSS 
cell (50-100\,Mpc), an even more distinct effect is expected here. 

Indeed, the line luminosity density varies from field to field by a factor 
of about 2 (ratio of highest to lowest values for identical redshift 
intervals). An average over only three fields leaves an uncertainty 
of the order 40\%.

\subsection{Emission line luminosities versus UV luminosities}

For optically selected galaxy samples the SFR can also be estimated by means 
of the UV emission. 
While the emission lines from H\,{\sc ii} regions measure the rate of massive 
stars born less than a few million years ago, the integrated UV flux from short-lived 
stars indicates the star formation rate of a somewhat older star population. 
For large redshifts ($z > 2.8$) where the observation of emission lines is 
extremely difficult and the L$\alpha$ line is often buried by internal dust 
extinction,  Steidel et al. (1999) used the UV flux at 300\,nm to derive star formation 
rates at high redshifts. 

The conversion factor between UV flux and SFR is still a matter of debate 
and is sensitive to the metallicity and IMF used in the model calculation 
(Glazebrook et al. 1999), but also to the intrinsic dust extinction adopted. 

We used for the extinction in the UV continuum the relation given by 
Calzetti et al. (1994): ~ 
${\tau_{\rm B}}^c = 0.5 {\tau_{\rm B}}^l$, \\
where ${\tau_{\rm B}}^c$ and ${\tau_{\rm B}}^l$ are the optical depths for starlight 
continuum emission and for line emission from the H\,{\sc ii} regions, respectively. 

For the redshifts discussed above, the 280\,nm (rest frame) continuum corresponds 
to observed wavelengths inbetween 350\,nm and 616\,nm. Except for the lowest 
redshift bin ($z=0.25$), the UV flux density at 280\,nm can thus be easily 
determined from the CADIS multi-filter data. In the case of $z=0.25$, we use 
the 396\,nm filter data if available, corresponding to a rest wavelength of 
316\,nm instead of 280\,nm. To account for this wavelength difference, a 
correction factor of (280/316)$^2$ (corresponding to a flat spectrum 
$F_{\nu}=constant$) is applied to the UV flux densities.

\begin{figure}[ht!]          % Fig 11
\centerline{\psfig{figure=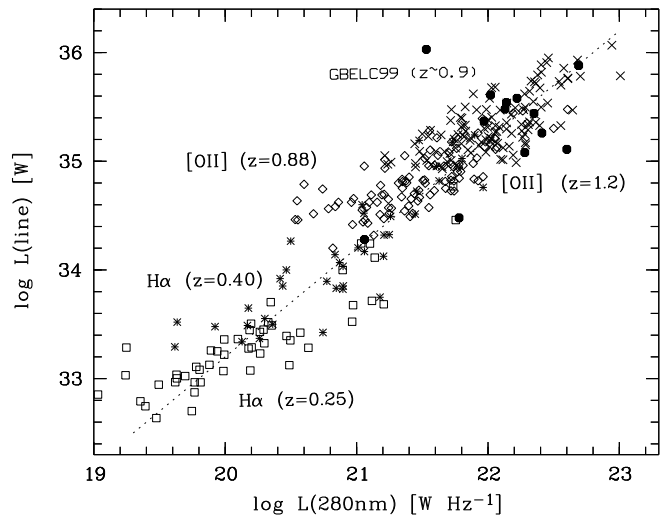,clip=t,width=8.8cm,angle=0}}
\caption[]{UV continuum flux density at 280\,nm plotted versus emission 
line strengths, both reddening corrected, for redshifts between 0.25 and 1.2. 
Only objects with UV flux signal-to-noise ratios above 2 are shown here. 
The dotted line is for a Bruzual \& Charlot model with solar abundance 
and Salpeter IMF and assuming the emission line galaxies are seen 0.1\,Gyr 
after the star burst (Glazebrook et al. 1999). }
\end{figure}

The relation between UV continuum at 280\,nm and emission line fluxes is shown 
in Fig.\,11. In this graph, the [O\,{\sc ii}] fluxes are scaled up by a factor 
1.1 to account for the average line intensity ratio between H$\alpha$ and [O\,{\sc ii}]. 
[O\,{\sc iii}] line fluxes, which show a large scatter due to varying metallicity 
and excitation for individual galaxies, are not included; in the case of 
$z=0.4$, we however used the corresponding H$\alpha$ luminosities, 
measured in the atmospheric window C at 918\,nm. 
The distribution shows no obvious deviation from proportionality over almost 
4 decades, and it follows rather close the predictions from the Bruzual-Charlot 
model for solar abundance. Thus, the UV\,280\,nm continuum data for the CADIS 
galaxies are expected to yield a similar SFR as the line flux densities. 

The CFRS galaxies at $z\sim0.9$ studied by Glazebrook et al. (1999) (filled 
circles in the diagram) agree with our data. Their finding that the mean star 
formation rate derived their from H$\alpha$ data is three times as high as the 
ultraviolet estimates seems to become obsolete when the reddening correction 
proposed in a footnote in their table 2 is applied (1.6 for H$\alpha$ and of 
3.1 for $L(280)$, almost identical with the corrections applied in the present 
paper and with the reddening correction proposed by Steidel et al., 1999). 
The remaining offset to higher luminosities of the Glazebrook galaxies as 
compared to our $z=0.88$ sample is due to the fact that the CFRS survey covers 
a larger volume than the present study, but is less deep in terms of emission 
line luminosities. 

Sullivan et al. (2000, 2001) did a similar comparison based on a UV selected 
sample of galaxies in a redshift range $0<z<0.4$. They also find agreement 
with the Glazebrook et al. (1999) data for the bright end of the sample. At the 
faint end, however, the UV fluxes are too high as compared to the H$\alpha$ 
line emission, not explainable by insufficient dust extinction correction. 
Sullivan et al. explain this discrepancy by series of starbursts superimposed 
on underlying quiescent star formation of the galaxies. Interestingly our 
emission line selected galaxy samples show an effect in the opposite direction 
(taking each redshift bin separately), which can be explained by exactly the 
same effect. Whereas an emission line selected galaxy sample prefers those 
(faint) galaxies which just happen to be in a starburst phase, a UV selected 
sample preferentially selects those which are in a state about 10$^8$ years past 
their starburst.

% Sullivan 2000 : SFR = 0.87e-34 LHalpha/W
%                      = 0.80e-34 L[OII]/W
%  Sullivan 2001: SFR = 0.92e-34 LHalpha/W
%                     = 0.18e-32 LUV/W/A = 0.18e-32 LUV/W (nue/lambda)/Hz 
%                     = 0.18e-32 LUV/W (3e14/0.280 /2800)/Hz
%                     = 1.9e-18/2800 = e-21.16 LUV/W/Ha   (for 100Myr)
%                                      e-21.0  LUV/W/Hz   (for 10Myr)

\subsection{Comparison with other studies}

In Fig.\,12, our values are inserted in a diagram similar to that presented 
by Glazebrook et al. (1999), with the lookback time as abscissae, showing the 
optical SFR values as well as those derived from FIR/sub-mm and UV luminosities. 
Here, all SFR values from UV studies are corrected by factors between 2.7 and 4.5, 
depending on the rest wavelength observed, but irrespective the galaxy brightnesses. 
This extinction correction is based on a global value $E_{\rm B-V}=0.15$, as 
proposed by Steidel et al. (1999). Note that this value corresponds to an 
emission line extinction correction with $E_{\rm B-V}=0.3$ (Calzetti et al. 1994), 
typical for a $M_{\rm B}=20$\,mag galaxy. 
The diagram shows that with this correction the different approaches to 
determine the SFR lead to satisfactorily consistent distributions 
(the error bars do not include systematic errors).

While our SFR values for redshift below 1.0 are close to those published 
by other authors, our value at $z=1.2$ seems to be significantly higher, 
probably due to the greater depth of the CADIS survey.

It is remarkable how closely the SFR data from infrared and sub-mm observations 
follow those derived from optical observations, although IRAS and ISO observations 
show that a notable part of the IR radiation density arises from galaxies barely 
visible in the optical, such as ULIRGs and AGNs. 
This agreement expresses the consistency of the conversion factors for the different 
kind of data, and the plausibility of the corrections to be applied. 
A close agreement between SFR densities derived from different wavelength regimes 
was also presented by Rosa-Gonzalez et al. (2002), who used global factors for 
the extinction correction of the luminosity densities, independent of redshift. 

The increase of the star forming rate with lookback time in Fig.\,12 can be well 
described by SFR$ \propto exp(t_{\rm lookback}/2.6)$ (dotted line), 
with the lookback time given in Gyrs. 
This exponential relation fits better than the power law $(1+z)^{3.9}$ fit 
proposed by Lilly et al. (1996) for the UV luminosity evolution.

\begin{figure}[h]                % Fig 12
\centerline{\psfig{figure=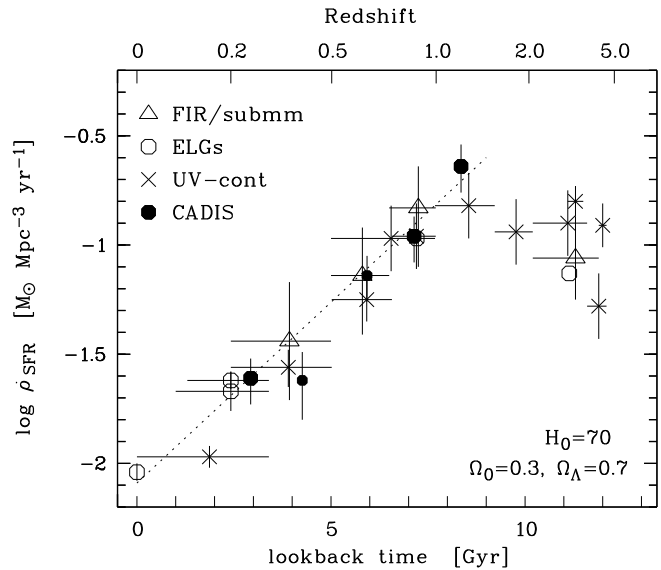,clip=t,width=8.8cm,angle=0}}
\caption[]{SFR density vs. redshift. CADIS points at $z=0.25$, 
0.4, 0.64, and 1.2 (black dots) are inserted in this diagram adopted from 
figure 8 of Glazebrook et al. (1999). 
UV data are from Sullivan et al. (2000), Lilly et al (1996), Connolly 
et al. (1997) and Madau et al. (1996), FIR and sub-mm data from Flores et al. 
(1999), and Hughes et al. (1998), emission line data from Gallego et al. 
(1995) at $z=0$, Tresse \& Maddox (1998), Glazebrook 
et al. (1999), and Pettini et al. (1998). 
UV values are extinction corrected according to Steidel et al. (1999). 
Literature SFR values are transformed to the cosmology used in the 
the present paper (parameters in the lower right of the figure). The dotted 
line represents a hand-drawn fit to the data. }
\end{figure}

While the UV fluxes are closely correlated to the emission line luminosities, 
irrespective the redshift (Fig.\,11), the flux density in the blue decreases 
by a factor of only $\sim4$ between $z=1$ and today, compatible with passive 
evolution (Fried et al. 2001). 
Comparing the $M_{\rm B}^{\ast}$ values from Fig.\,5 with the corresponding $L^{\ast}$ 
values marked in Fig.\,6 (both Schechter functions have the identical $\phi$ and 
$\alpha$ parameters), one can see, that the ratio between blue flux density 
and emission line flux density (star formation rate) increases by a factor $\sim5$ 
between $z=0.25$ and 1.2. Thus, the star forming rate per blue luminosity unit 
increases drastically with time. The slower evolution of the blue luminosity 
density indicates that the relation between moderately young stars (age 
$\simeq 2$\,Gyr) to very young stars (age $\la 0.1$\,Gyr) evolves with time 
in that sense that a higher redshift starbursts were more frequent than 
today. 

Two interpretations are possible for the exponential decay of the SFR. One is 
that on average the amount of gas available 
in the galaxies determines the SFR and thus young galaxies have larger star 
forming activities. The other interpretation is that independent of the gas 
content, a high star forming activity is induced by galaxy-galaxy interactions. 
In this case the interpretation would be that more interactions must have occurred 
at high redshifts. Both of these could have an exponential decay with time. 
The latter interpretation is supported by LeFevre et al. (2000) who find an 
increase of the interaction rate with $(1+z)^{2.7}$, resulting in a factor 8 
between today and $z=1.2$. 
Model calculations show that an exponential law fits perfectly to the 
merger rates (Khochfar et al. 2003).

\section{Conclusions}

Based on the photometric emission line data the star formation rates for 
redshifts ranging from 0.25 and 1.2 are derived. 
This CADIS emission line survey used for the present study survey is deeper 
than similar studies performed up to now and thus the resulting luminosity 
densities and star formation rates are less affected by the completeness 
correction of the luminosity functions. All results, including broad band 
colors, are based on only one data set. 

The SFR increases by $\sim20$ from $z=0$ to $z=1.2$, following an exponential 
relation  ${\rm SFR} \propto exp(t_{\rm lookback}/2.6)$. 

The (blue) luminosity functions derived from the line emission galaxy samples 
agree well with those derived from the multi-color survey of CADIS at 
comparable redshifts. Also, the emission line luminosity functions agree in 
shape ($\Phi^{\ast}$ and $\alpha$) with the corresponding blue magnitude luminosity 
functions, although $L^{\ast}_{\rm lines}$ evolves $\sim5$ times faster than 
$M_{\rm B}^{\ast}$ within the studied redshift range of $z=0.25$ to 1.2. 

A steep rise of SFR with lookback time up to 8\,Gyrs is clearly indicated, 
and can well described with an exponential decay of SFR with time, with a 
(half) decay time of $\sim2.6$\,Gyrs. 

The extinction correction plays an important role in the determination of 
the SFR and leads to major (systematic) uncertainties; for higher 
redshifts, where the Balmer line ratio cannot be determined accurate enough, 
the correction can be only carried out with global methods. A reddening 
correction depending on absolute magnitude leads to consistent results.

\acknowledgements{
We thank A. Aguirre and M. Alises from the Calar Alto Observatory for carefully 
carrying out observations in service mode. We also wish to thank Drs. Henry Lee, 
Eric Bell, Andreas Burkert and Sadegh Khochfar for valuable discussions.}


\begin{thebibliography}{}
\bibitem[]{} Baldwin, J.A., Phillips, M.M., \& Terlevich, R.  1981, PASP, 93, 5
\bibitem[]{} Bell, E.F., \& Kennicutt, R.C.  2001, ApJ, 548, 681
\bibitem[]{} Bertin, E., \& Arnouts, S,  1996, A\&A, 117, 393
\bibitem[]{} Carollo, C.M., \& Lilly, S.J. 2001, ApJ, 548, L53
\bibitem[]{} Connolly, A.J., Szalay, A.S., Dickinson, M., Subbarao, M.U., \& 
             Brunner, R.J. 1997, ApJ, 486, L11
\bibitem[]{} Cowie, L.L., Hu, E.M., Songaila, A., \& Egami, E. 1997, ApJ, 481, L9
\bibitem[]{} Calzetti, D., Kinney, A.L., \& Storchi-Bergmann, T. 1994, ApJ, 429, 582
\bibitem[]{} Flores, H., Hammer, F., T. Thuan, T., et al.  1999, ApJ, 517, 148
\bibitem[]{} French, H. 1980, ApJ, 240, 41
\bibitem[]{} Fried, J.W., von Kuhlmann, B., Meisenheimer, et al.  2001, A\&A, 367, 788
\bibitem[]{} Gallagher, J.S., Bushouse, H., \& Hunter, D.A.  1989, AJ, 97, 700 
\bibitem[]{} Gallego, J., Zamorano, J., Aragon-Salamanca, A., \& Rego, M. 
           1995, ApJ, 455, L1
\bibitem[]{} Gallego, J., Zamorano, J., Rego, M., Alonso, O., \& Vitores, A.G. 
           1996, A\&AS, 120, 323
\bibitem[]{} Gallego, J., Zamorano, J., Rego, M., \& Vitores, A.G. 
           1997, ApJ, 475, 502
\bibitem[]{} Glazebrook, K., Blake, C., Economou, F., Lilly, S., \& Colless, M. 
           1999, MNRAS, 306, 843 
\bibitem[]{} Hammer, F., Flores, H., Lilly, S.J., et al.  1997, ApJ, 481, 49
\bibitem[]{} Ho, L.C., Filippenko, A.V., \& Sargent, W.L.W.  1997, ApJS, 112, 315
\bibitem[]{} Hogg, D.W. Cohen, J.G., Blandford, R., \& Phare, M.A.  1998, ApJ, 504, 622
\bibitem[]{} Hughes, D.H., Serjeant, S., Dunlop, J., et al.  1998, Nature, 394, 241 
\bibitem[]{} Jansen, R.A., Franx, M., \& Fabricant, D. 2001, ApJ, 551, 825
\bibitem[]{} Kennicutt, R.C. 1983, ApJ, 272, 54 
\bibitem[]{} Kennicutt, R.C. 1992, ApJ, 388, 310
\bibitem[]{} Kennicutt, R.C. 1998, ARA\&A, 36, 189
\bibitem[]{} Khochfar, S., et al. 2003, in preparation
\bibitem[]{} Kinney, A.L., Calzetti, D., Bohlin, R., et al. 1996, ApJ, 467, 38
\bibitem[]{} LeFevre, O., Abraham, R., Lilly, S., L., et al.  2000, MNRAS, 311, 565 
\bibitem[]{} Lilly, S.J., Le\,F\`evre, Hammer, F.O., \& Crampton. D. 1996, ApJ, 460, L1
\bibitem[]{} Madau, P., Ferguson, H.C., Dickinson, M., et al.  1996, MNRAS, 283, 1388
\bibitem[]{} Maier C., Meisenheimer, K., Hippelein, H., et al.  2003, submitted
\bibitem[]{} McCall, M.L., Rybski, P.M., \& Shields, G.A. 1985, ApJS, 57, 1
\bibitem[]{} Meisenheimer, K., \& R\"oser, H.-J.  1993, in Landolt-B\"ornstein, 
           Extension and Supplement to Volume 2, Subvolume a, 29 (Springer Verlag)
\bibitem[]{} Meisenheimer, K., Fried, J.W., Hippelein, H., et al.  2002, in preparation
\bibitem[]{} Pettini, M., Kellogg, M. Steidel, C.C., et al.  1998, ApJ, 508, 539
\bibitem[]{} Pettini, M., Shapley, A.E., Steidel, C.C., et al.  2001, ApJ, 554, 981
\bibitem[]{} Popescu, C.C., \& Hopp, U. 2000, A\&AS, 142, 247
\bibitem[]{} Rosa-Gonzalez, Terlevich, E., \& Terlevich, R.  2002, MNRAS, 332, 283
\bibitem[]{} Rowan-Robinson, M., Mann, R.G., Oliver, S.J., et al. 1997, MNRAS, 289, 490
\bibitem[]{} Steidel, C.C., Adelberger, K.L., Giavalisco, M., et al.  1999, ApJ, 519, 1
\bibitem[]{} Szokoly, G., et al., 2002, A\&A, submitted
\bibitem[]{} Sullivan, M., Treyer, M.A., Ellis, R.S., et al.  2000, MNRAS, 312, 442
\bibitem[]{} Sullivan, M., Mobasher, B., Chan, B., et al.  2001, ApJ, 558, 72
\bibitem[]{} Teplitz, H.I., Malkan, M.A., Steidel, C.C., et al.  2000, ApJ, 542, 18
\bibitem[]{} Terlevich, R., Melnick, J., Masegosa, J., et al.  1991, A\&AS, 91, 285
\bibitem[]{} Tresse, L., \& Maddox, S.J.  1998, ApJ, 495, 691
\bibitem[]{} Veilleux, S., \& Osterbrock, D.E. 1987, ApJS, 63, 295
\bibitem[]{} Vogel, S., Engels, D., Hagen, H.-J., et al. 1993, A\&AS, 98, 193
\bibitem[]{} Wolf, C., Meisenheimer, K., R\"oser, H.-J., et al.  2001, A\&A, 365, 681
\bibitem[]{} Yan, L., McCarthy, P.J., Freudling, W., et al.  1999, ApJ, 519, L47

\end{thebibliography}
\end{document}